\documentclass[a4paper,11pt]{article}
\pdfoutput=1 

\usepackage{jheppub} 

\begin{document} 

\title{\boldmath Measurement of the $K_S \to \pi e \nu$ branching fraction
with the KLOE experiment~$^*$}


\author[]{{\large The KLOE-2 Collaboration \vskip 0.2cm}}
\author[c]{D.~Babusci}
\author[p]{M.~Berlowski}
\author[c]{C.~Bloise}
\author[c]{F.~Bossi}
\author[n]{P.~Branchini}
\author[o]{B.~Cao}
\author[m,n]{F.~Ceradini}
\author[c]{P.~Ciambrone}
\author[h,i]{F.~Curciarello}
\author[b]{E.~Czerwi\'nski}
\author[k,l]{G.~D'Agostini}
\author[k,l]{R.~D'Amico}
\author[c]{E.~Dan\`e}
\author[k,l]{V.~De~Leo}
\author[c]{E.~De~Lucia}
\author[c]{A.~De~Santis}
\author[c]{P.~De~Simone}
\author[c]{A.~Di~Cicco}
\author[k,l]{A.~Di~Domenico}
\author[c]{E.~Diociaiuti}
\author[c]{D.~Domenici}
\author[c]{A.~D'Uffizi}
\author[k,l]{G.~Fantini}
\author[b]{A.~Gajos}
\author[b]{S.~Gamrat}
\author[k,l]{P.~Gauzzi}
\author[c]{S.~Giovannella}
\author[n]{E.~Graziani}
\author[q]{X.~Kang}
\author[o,p]{A.~Kupsc}
\author[e,a]{G.~Mandaglio}
\author[c,j]{M.~Martini}
\author[c]{S.~Miscetti}
\author[b]{P.~Moskal}
\author[n]{A.~Passeri}
\author[b]{E.~Perez~del~Rio}
\author[h,i]{M.~Schioppa}
\author[m,n,1]{A.~Selce}
\author[b]{M.~Silarski}
\author[c,d]{F.~Sirghi}
\author[f,g]{E.~P.~Solodov}
\author[p]{W.~Wi\'slicki}
\author[o]{M.~Wolke}
\affiliation[a]{NFN Sezione di Catania, Catania, Italy.}
\affiliation[b]{Institute of Physics, Jagiellonian University, Cracow, Poland.}
\affiliation[c]{Laboratori Nazionali di Frascati dell'INFN, Frascati, Italy.}
\affiliation[d]{Horia Hulubei National Institute of Physics and Nuclear Engineering, M\v{a}gurele, Romania.}
\affiliation[e]{Dipartimento di Scienze Matematiche e Informatiche, Scienze Fisiche e Scienze della Terra dell'Universit\`a di Messina, Messina, Italy.}
%
\affiliation[f]{Budker Institute of Nuclear Physics, Novosibirsk, Russia.}
\affiliation[g]{Novosibirsk State University, Novosibirsk, Russia.}
\affiliation[h]{Dipartimento di Fisica dell’Universit\`a della Calabria, Arcavacata di Rende, Italy.}
\affiliation[i]{INFN Gruppo collegato di Cosenza, Arcavacata di Rende, Italy.}
%
\affiliation[j]{Dipartimento di Scienze e Tecnologie applicate, Universit\`a Guglielmo Marconi, Roma, Italy.}
\affiliation[k]{Dipartimento di Fisica dell’Universit\`a Sapienza, Roma, Italy.}
\affiliation[l]{NFN Sezione di Roma, Roma, Italy.}
\affiliation[m]{Dipartimento di Matematica e Fisica dell'Universit\`a Roma Tre, Roma, Italy.}
\affiliation[n]{INFN Sezione di Roma Tre, Roma, Italy.}
\affiliation[o]{Department of Physics and Astronomy, Uppsala University, Uppsala, Sweden.}
\affiliation[p]{National Centre for Nuclear Research, Warsaw, Poland.}
\affiliation[q]{School of Mathematics and Physics, China University of Geosciences, Wuhan, China.}
\affiliation[1]{Corresponding author {\bf andrea.selce@roma3.infn.it} \\}
\affiliation[*]{{\bf \ Dedicated to the memory of Paolo Franzini}}

\abstract{
The ratio $\mathcal{R} = \Gamma(K_S \to \pi e \nu)/\Gamma(K_S \to \pi^+ \pi^-)$ has been measured with a sample
of 300 million $K_S$ mesons produced in $\phi \to K_L K_S$ decays recorded by the KLOE
experiment at the DA$\Phi$NE $e^+e^-$ collider. 
$K_S \to \pi e \nu$ events are selected by a boosted
decision tree built with kinematic variables and time-of-flight measurements. Data
control samples of $K_L \to \pi e \nu$ decays are used to evaluate signal selection efficiencies.
With 49647$\pm$316 signal events we measure $\mathcal{R} = (1.0421 \pm 0.0066_{\rm stat} \pm 0.0075_{\rm syst})\times10^{-3}$.
The combination with our previous measurement gives 
$\mathcal{R} = (1.0338 \pm 0.0054_{\rm stat} \pm 0.0064_{\rm syst})\times10^{-3}$.
From this value we derive the branching fraction 
$\mathcal{B}(K_S \to \pi e \nu) = (7.153 \pm 0.037_{\rm stat} \pm 0.044_{\rm syst}) \times10^{-4}$ and 
$f_+(0)|V_{us}| = 0.2170 \pm 0.009$.
}

\maketitle
\flushbottom

\section{Introduction} \label{INTRO}
The branching fraction for semileptonic decays of charged and neutral kaons
together with the lifetime measurements are used to determine the $|V_{us}|$ 
element of the Cabibbo--Kobayashi--Maskawa 
quark mixing matrix. The relation among the matrix elements of the first row, 
$|V_{ud}|^2 + |V_{us}|^2 + |V_{ub}|^2 = 1$, 
provides the most stringent test of the unitarity of the quark mixing matrix.
At present, the sum $\sum_i |V_{ui}|^2$ differs from one by about $3\sigma$, 
an intriguing question under careful scrutiny,
the so-called Cabibbo angle anomaly~\cite{Grossman2020}.

Different factors contribute to the uncertainty in determining $|V_{us}|$ from kaon
decays, discussed in 
Refs.~\cite{ref:Antonelli2010, ref:Passemar2018, ref:PDGVus, ref:VusUpdate},
and among the six semileptonic 
decays the contribution of the lifetime uncertainty is smallest 
for the $K_S$ meson. Nevertheless, given the lack of pure high-intensity
$K_S$ meson beams compared with $K^{\pm}$ and $K_L$ mesons, 
the measurements of $K_S$ semileptonic decays  
provide the least precise 
determination of $|V_{us}|$.
Beside early measurements of $\mathcal{B}(K_S \to \pi e \nu)$ based on 
$\phi \to K_L K_S$ decays~\cite{ref:CMD2, ref:KLOE2002} and the recent measurement of 
$\mathcal{B}(K_S \to \pi \mu \nu)$~\cite{ref:KStopimunu},
the most precise measurements of the $K_S$ semileptonic
branching fraction are from NA48: $\mathcal{B}(K_S \to \pi e \nu) =
(7.05 \pm 0.18_{\rm stat} \pm 0.16_{\rm syst})\times10^{-4}$~\cite{ref:Na48}, and KLOE: $\mathcal{B}(K_S \to \pi e \nu) =
(7.046 \pm 0.091)\times10^{-4}$~\cite{ref:KStopienu}.

We present a new measurement of the ratio $\mathcal{R} =
\frac{\Gamma(K_S \to \pi e \nu)}{\Gamma(K_S \to \pi^+ \pi^-)}$
performed by the KLOE experiment at the
DA$\Phi$NE $\phi$--factory of the Frascati National Laboratory
based on data collected in 2004--05 corresponding to
an integrated luminosity of 1.63 fb$^{-1}$.
DA$\Phi$NE~\cite{ref:DAFNE} is an electron--positron collider
running at the centre-of-mass energy of $\sim$1.02 GeV colliding $e^+$ and $e^-$
beams at an angle of $\pi$$-$0.025 rad and with a bunch-crossing period of 2.715 ns. The $\phi$ mesons are
produced with a small transverse momentum $p_{\phi}$ of $\sim$13 MeV and $K_L$--$K_S$ pairs are produced almost back-to-back with an effective
cross section of $\sim$1 mb.
The beam energy, the energy spread, the $\phi$
transverse momentum and the position of the interaction point
are measured with high accuracy using Bhabha scattering 
events~\cite{ref:Luminosity}.

The $K_S$ ($K_L$) mesons are identified (\textit{tagged}) with 
high efficiency and purity by the observation of a $K_L$ ($K_S$) 
in the opposite hemisphere.
This tagging procedure allows the selection
efficiency for $K_S \to \pi e \nu$ to be evaluated with good 
accuracy using a sample of the abundant decay $K_L \to \pi e \nu$ 
tagged by the detection of $K_S \to \pi^+ \pi^-$ decays. 
The branching fraction $\mathcal{B}(K_S \to \pi e \nu)$
is obtained from the ratio $\mathcal{R}$ and the value of 
$\mathcal{B}(K_S \to \pi^+ \pi^-)$ measured by KLOE~\cite{ref:KStopipi}.

\section{The KLOE detector} \label{DETECTOR}
The detector consists of a large-volume cylindrical drift chamber, 
surrounded by a lead-scintillating fibers finely-segmented calorimeter. 
A superconducting coil around the calorimeter provides a 0.52 T axial 
magnetic field. The beam pipe at the interaction region
is spherical in shape with 10 cm radius, made of a 0.5 mm thick 
beryllium--aluminium alloy. Low-beta quadrupoles are located at $\pm$50 cm 
from the interaction region. Two small lead-scintillating-tile 
calorimeters~\cite{ref:QCAL} are wrapped around the quadrupoles.

The drift chamber (DC)~\cite{ref:DC}, 4 m in diameter and 3.3 m long, has 
12582 drift cells arranged in 58 concentric rings with alternated stereo angles 
and is filled with a low-density gas mixture of 90\% helium--10\% isobutane. 
The chamber shell is made of carbon fiber-epoxy composite with an 
internal wall of 1.1 mm thickness at 25 cm radius. The spatial
resolution is $\sigma_{xy} =$ 0.15 mm and $\sigma_z =$ 2 mm
in the transverse and longitudinal projection, respectively. 
The momentum resolution for tracks with polar angle 
$45^{\circ} < \theta < 135^{\circ}$ is $\sigma_{p_{\rm T}}/p_{\rm T} = 0.4\%$. 
Vertices formed by two tracks are reconstructed with a spatial resolution of about 3 mm.

The calorimeter (EMC)~\cite{ref:EMC} is divided into a barrel and two endcaps 
and covers 98\% of the solid angle. The readout granularity is 
4.4$\times$4.4 cm$^2$, for a total of 2440 cells arranged in
five layers. Each cell is read out at both ends by photomultipliers. 
The energy deposits are obtained from signal amplitudes, 
the arrival times of particles and their position along the fibres are determined from the signals
at the two ends. 
Cells close in space and time are grouped into energy clusters. The cluster energy $E$ is the sum of the cell energies, 
the cluster time and position are energy-weighted averages. 
Energy and time resolutions are $\sigma_E/E = 0.057/\sqrt{E\ {\rm (GeV)}}$
and $\sigma_t = 54\ {\rm ps}/\sqrt{E\ {\rm (GeV)}} \oplus 100$ ps, respectively. 
The cluster spatial resolution is 
$\sigma_{\parallel} = 1.4\ {\rm cm}/\sqrt{E\ {\rm (GeV)}}$
along the fibres and $\sigma_{\perp} = 1.3$ cm in the orthogonal direction.

The level-1 trigger~\cite{ref:Trigger} uses both the calorimeter and the drift chamber 
information; the calorimeter trigger requires two energy deposits 
with $E > 50$ MeV in the barrel and $E > 150$ MeV in the endcaps;
the drift chamber trigger is based on the number and topology of 
hit drift cells. A higher-level cosmic-ray veto rejects events with at least two energy deposits above 30 MeV in the outermost calorimeter
layer. The trigger time is determined by the first particle reaching
the calorimeter and is synchronised with the DA$\Phi$NE r.f. signal.
The time interval between bunch crossings is smaller than the time
spread of the signals produced by the particles, thus the event T$_0$
related to the bunch crossing originating the event is determined after
event reconstruction and all the times related to that event are 
shifted accordingly. 
Data for reconstruction are selected by an on-line filter~\cite{ref:Datarec} to reject beam backgrounds. The filter also streams the events into 
different output files for analysis according to their properties and
topology. A fraction of 5\% of the events are recorded 
without applying the filter to control inefficiencies in the event streaming. 

The KLOE Monte Carlo (MC) simulation package, \texttt{GEANFI}~\cite{ref:Datarec}, has been used to produce an event sample equivalent to the data. 
Energy deposits in EMC and DC hits from beam background events triggered 
at random are overlaid onto the simulated events which are then processed with the same reconstruction algorithms as the data.

\section{ The measurement of 
$\Gamma(K_S \to \pi e \nu)/ \Gamma(K_S \to \pi^+ \pi^-)$} \label{BRMEAS}
The ratio of $\Gamma(K_S \to \pi e \nu)$ to $\Gamma(K_S \to \pi^+ \pi^-)$ is 
evaluated as
\begin{equation}
\mathcal{R} = \frac{\Gamma(K_S \to \pi e \nu)}{\Gamma(K_S \to \pi^+ \pi^-)} =  
\frac{N_{\pi e \nu}}{\epsilon_{\pi e \nu}} \times
\frac{\epsilon_{\pi \pi}}{N_{\pi \pi}} \times R_{\epsilon} ,
\label{eq:RATIO}
\end{equation}
where $N_{\pi e \nu}$ and $N_{\pi \pi}$ are the numbers of 
selected $K_S \to \pi e \nu$ and $K_S \to \pi^+ \pi^-$ events, 
$\epsilon_{\pi e \nu}$ and $\epsilon_{\pi \pi}$
are the respective selection efficiencies, and 
$R_{\epsilon} = (\epsilon_{\pi\pi}/\epsilon_{\pi e \nu})_{\rm com}$ is the
ratio of common efficiencies for the trigger, on-line filter, event classification and preselection that can be different for the two decays.

The number of signal events, $N_{\pi e \nu}$ in Eq. (\ref{eq:RATIO}), 
is the sum of the two charge-conjugated
decays to $\pi^- e^+ \nu$ and $\pi^+ e^- \bar{\nu}$. These are separated 
in a parallel analysis of the same dataset
based on the same selection criteria presented in this section, 
optimised for measuring the charge asymmetry 
$\frac{\Gamma(\pi^- e^+ \nu) - \Gamma(\pi^+ e^- \bar{\nu})}
         {\Gamma(\pi^- e^+ \nu) + \Gamma(\pi^+ e^- \bar{\nu})}$~\cite{ref:KStopienuAsimmetry}.

\subsection{Data sample and event preselection} \label{DATASAMPLE}
Neutral kaons from $\phi$-meson decays are emitted in two opposite hemispheres
with $\lambda_S = 5.9$ mm and $\lambda_L = 3.4$ m mean decay path for $K_S$ and $K_L$ respectively. About 50\% of $K_L$ mesons reach the calorimeter
before decaying and the $K_L$ velocity in the $\phi$-meson reference system is $\beta^* = 0.22$.
$K_S$ mesons are tagged by $K_L$ interactions 
in the calorimeter, $K_L$--crash in the following, with a clear signature
of a delayed cluster not associated to tracks.
To select $K_L$--crash and then tag $K_S$ mesons, the requirements are:
\begin{itemize}
\item one cluster not associated to tracks (neutral cluster) and with energy $E_{\rm clu} > 100$ MeV, the centroid of the neutral cluster defining the $K_L$ 
direction with an angular resolution of $\sim$1$^{\circ}$; 
\item $15^{\circ} < \theta_{\rm clu} < 165^{\circ}$ for the polar angle of the neutral cluster, to suppress small-angle beam backgrounds;
\item $0.17 < \beta^* < 0.28$ for the velocity in the $\phi$ reference 
system of the $K_L$ candidate; $\beta^*$ is obtained from 
the velocity in the laboratory system, 
$\beta = r_{\rm clu} / c t_{\rm clu}$, with $t_{\rm clu}$ being the cluster time 
and $r_{\rm clu}$ the distance from the nominal interaction point,
the $\phi$-meson momentum and the angle between the $\phi$-meson momentum and 
the $K_L$--crash direction.
\end{itemize}
The $K_S$ momentum 
$\vec{p}_{K_S} = \vec{p}_{\phi} - \vec{p}_{K_L}$ is determined 
with an accuracy of 2 MeV, assigning the neutral kaon mass. 

$K_S \to \pi e \nu$ and $K_S \to \pi^+ \pi^-$ candidates are preselected
requiring two tracks of opposite curvature forming a vertex inside the cylinder
defined by 
\begin{equation}
\rho_{\rm vtx} = \sqrt{x^2_{\rm vtx}  + y^2_{\rm vtx}} < 5\ {\rm cm}  
\qquad |z_{\rm vtx}| < 10\ {\rm cm} .
\label{eq:Vertex}
\end{equation}

After preselection, the data sample 
contains about 300 million events and its composition evaluated by simulation 
is shown in Table~\ref{tab:EventPre}.
The large majority of events are $K_S \to \pi^+ \pi^-$ decays, together with a large contribution from $\phi \to K^+ K^-$ events where one kaon produces a track and the kaon itself or its decay products generate a fake $K_L$--crash while the other kaon 
decays early into $\pi^{\pm}\pi^0$.
\begin{table}[htp!]
\caption{Number of events for data and simulation after $K_L$--crash  and $K_S$ preselection.}
\begin{center}
\begin{tabular}{lrr}
& Events & Fraction [\%] \\ 
\hline
Data &301\ 645\ 500 \\
MC &312\ 018\ 500\\
\hline
$\phi \to K_L K_S, K_S \to \pi e \nu$ & 259\ 264 & 0.08 \ \ \\
$\phi \to K_L K_S, K_S \to \pi^+\pi^-$ & 301\ 976\ 400&  96.78 \ \ \\
$\phi \to K^+K^-$ & 9\ 565\ 465 & 3.07 \ \ \\
$\phi \to K_L K_S, K_S \to \pi \mu \nu$ &139\ 585 &    0.04 \ \   \\
$\phi \to K_L K_S, K_S \to \pi^0\pi^0$ & 30\ 353&  0.01 \ \  \\
$\phi \to K_L K_S, K_S \to \pi^+\pi^- e^+ e^-$ & 18\ 397 & \  0.006  \\
$\phi \to \pi^+ \pi^- \pi^0$ & 24\ 153 & \     0.008 \\
others & 4\ 852 & \ 0.002 \\
\hline
\end{tabular}
\end{center}
\label{tab:EventPre}
\end{table}
 
\begin{figure}[htb!]
 \centering
  \includegraphics[width = 0.55\textwidth]{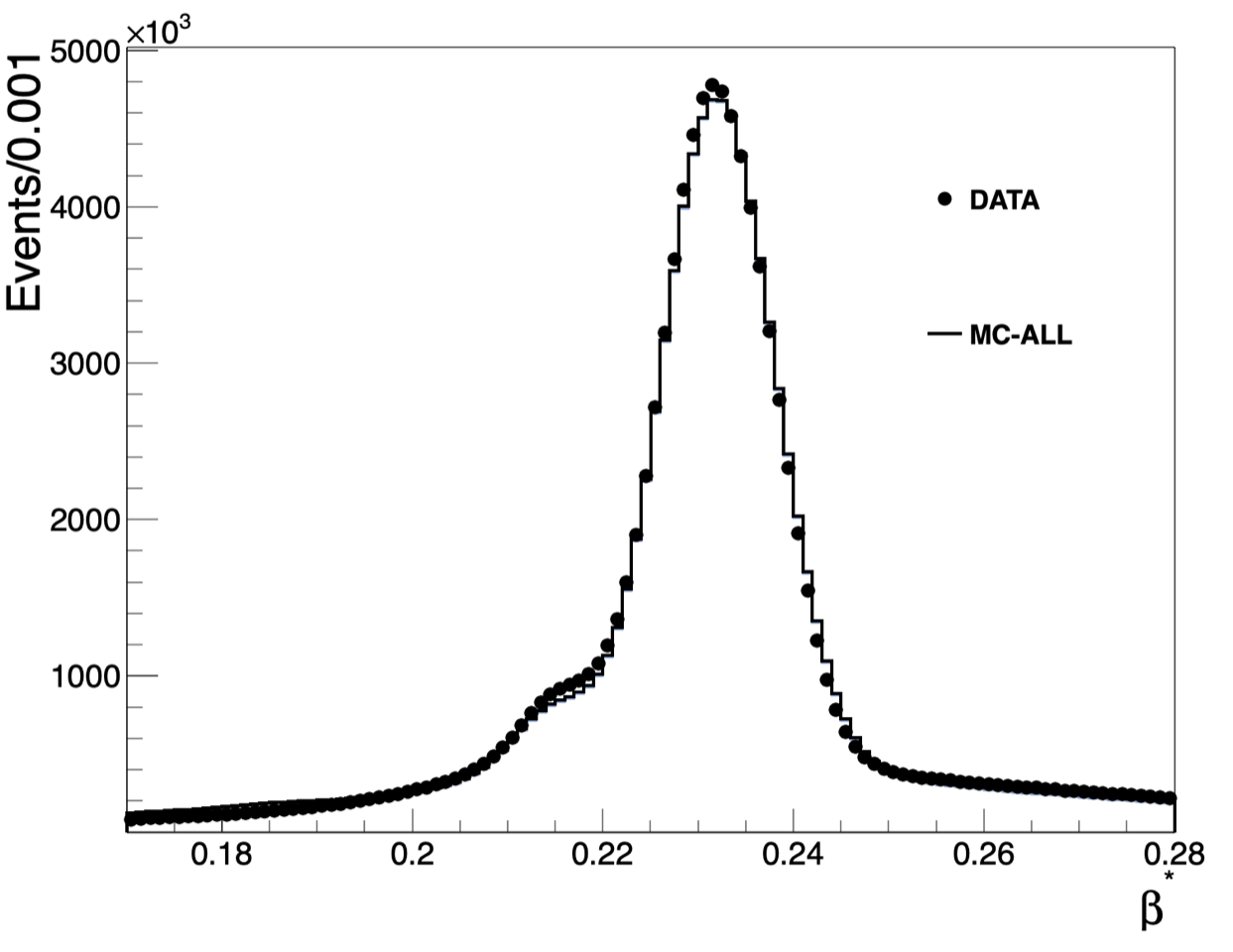}
 \caption{Distribution of $\beta^*$ after preselection for data and simulated events.}
 \label{fig:Beta}
\end{figure}

The $\beta^*$ distribution is shown in Figure~\ref{fig:Beta},
for data and simulated events. 
Two peaks are visible, the first is associated to events triggered by 
photons or electrons, and the second to events triggered
by charged pions. The trigger is synchronised with the bunch crossing and
the time difference between an electron (or photon) and a pion (or muon)
arriving at the calorimeter corresponds to about one bunch-crossing shift.

\subsection{Signal selection and normalisation sample} \label{KtoPIENU}
Signal selection is performed in two steps based on uncorrelated information: 1) the event kinematics using only DC tracking variables, and 2) the time-of-flight measured with the EMC. 

Time assignment to tracks requires track-to-cluster association 
(TCA): for each track connected to the vertex a cluster with 
$E_{\rm clu} > 20$ MeV and $15^{\circ} < \theta_{\rm clu} < 165^{\circ}$ 
is required whose centroid is within 30 cm of the track extrapolation inside the 
calorimeter. Track-to-cluster association is required for both tracks in the event.

\begin{figure}[htb!]
 \centering
 \begin{tabular}{@{}cc@{}}
 \includegraphics[width = 6.cm]{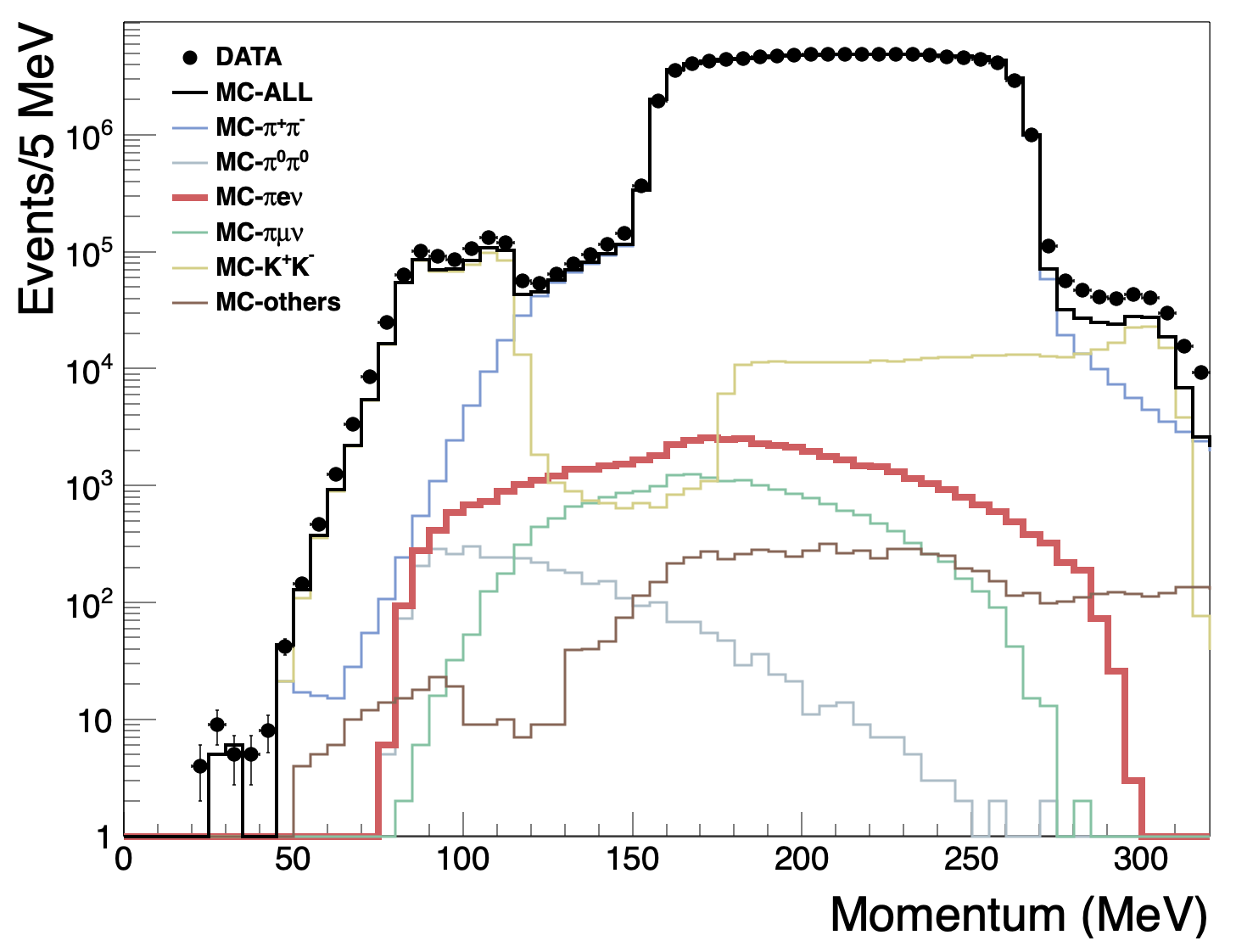}&
 \includegraphics[width = 6.cm]{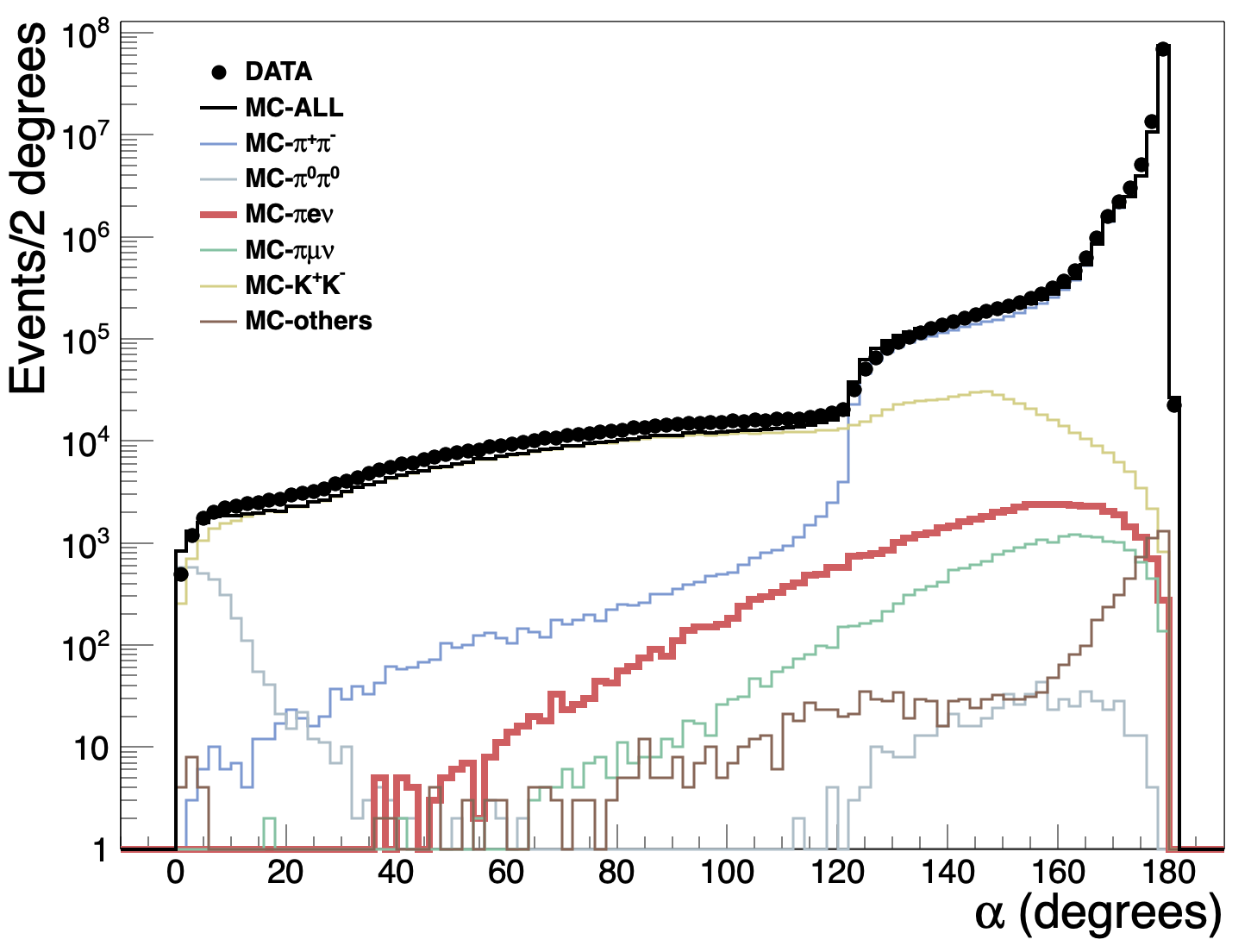}\\
 \includegraphics[width = 6.cm]{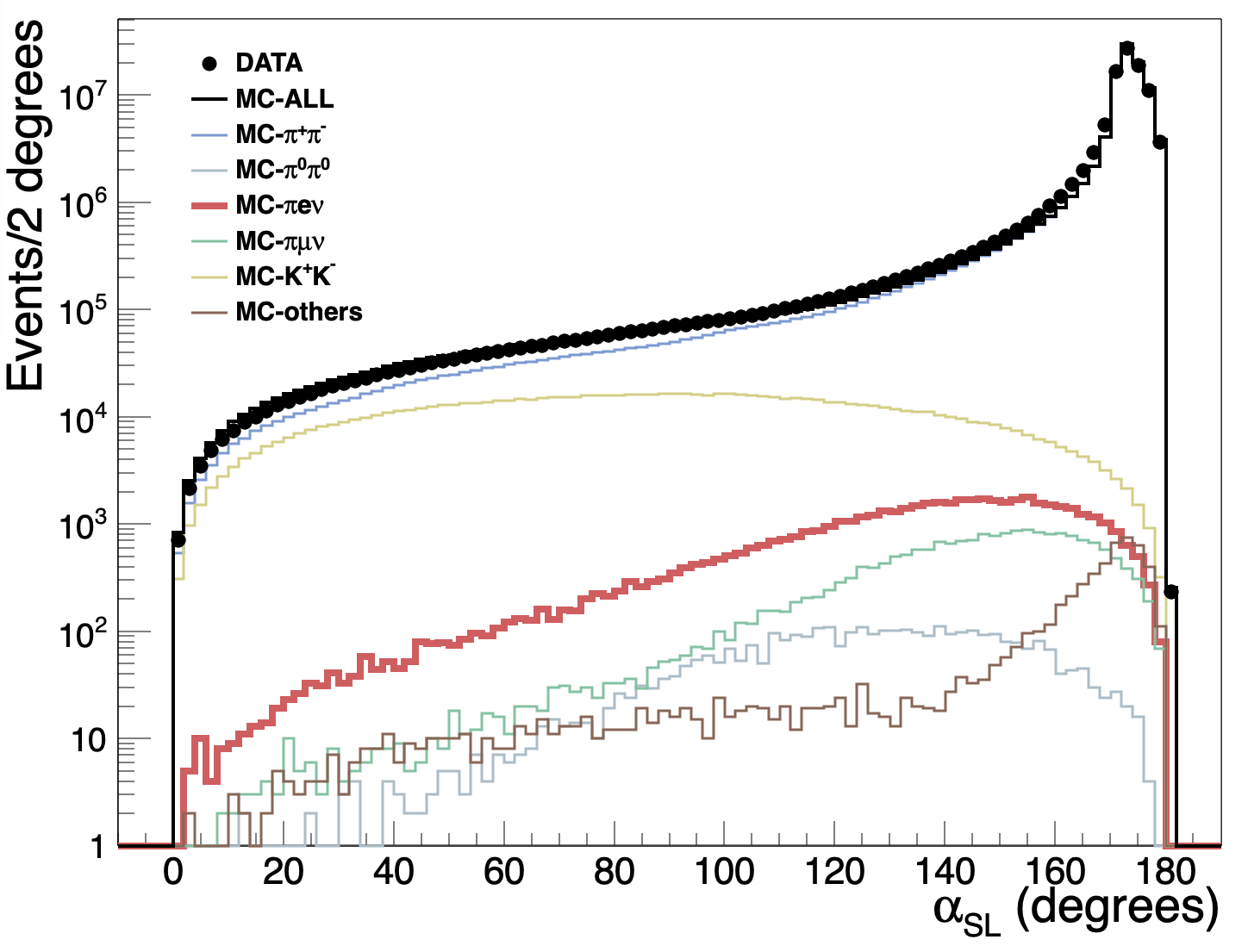}&
 \includegraphics[width = 6.cm]{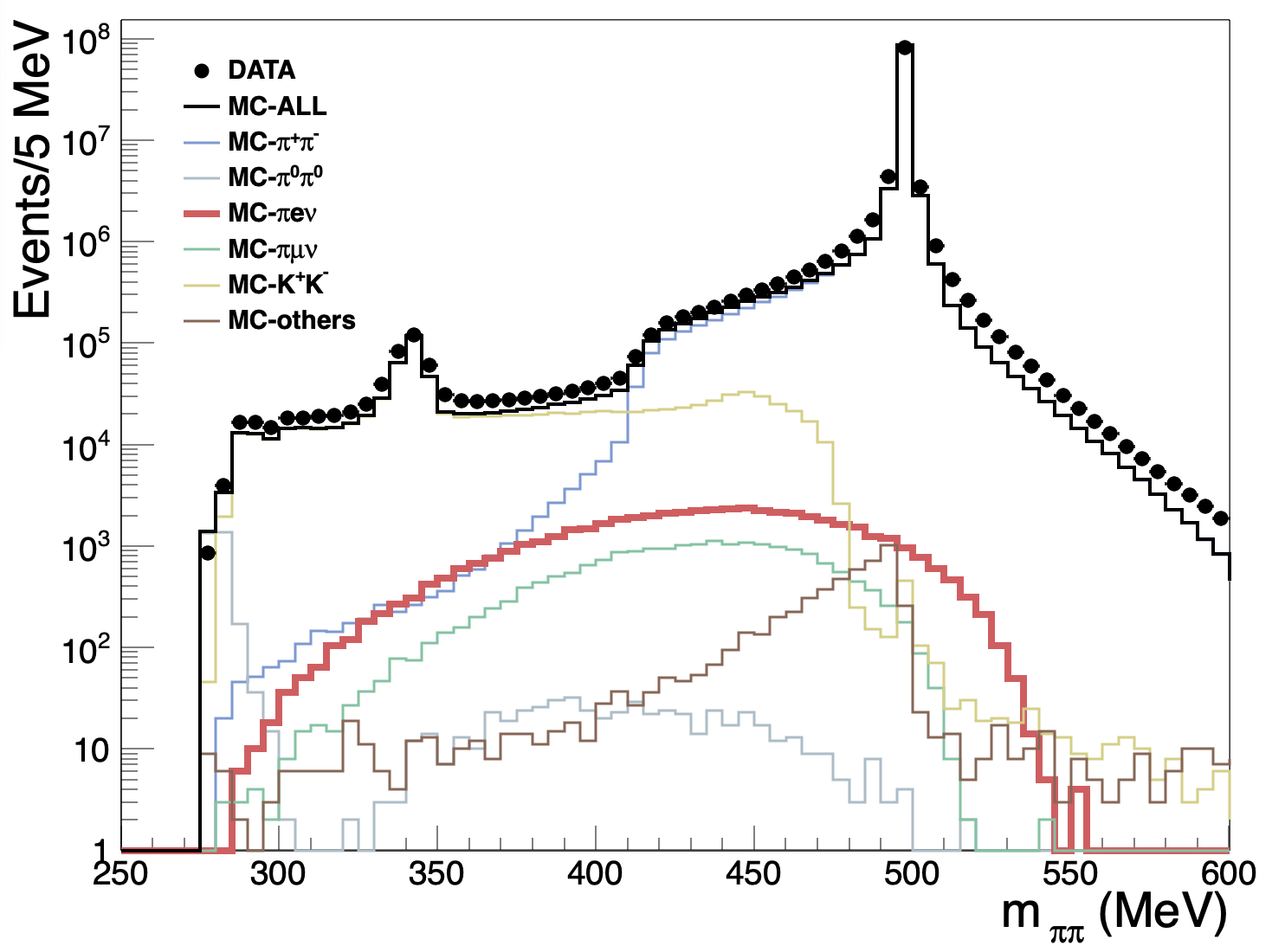}\\
 \includegraphics[width = 6.cm]{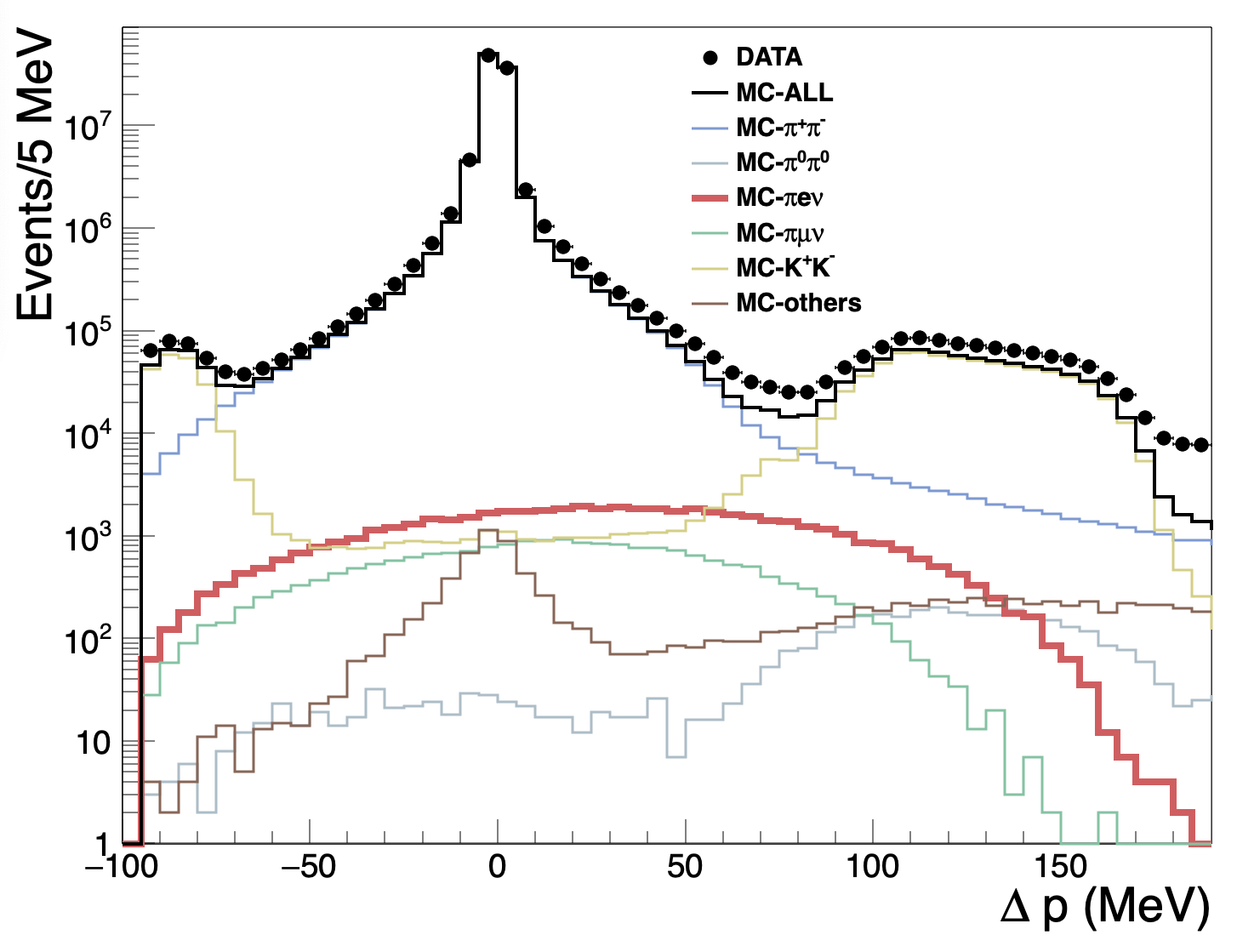}\\
  \end{tabular}
 \caption{Distributions of the variables used in the multivariate analysis for 
 data and simulated events after preselection.
 From top left: 
 track momenta ($p_1, p_2$),
angle between the two tracks in the $K_S$ reference system ($\alpha_{1,2}$), 
angle beween $K_L$ and $K_S$ directions ($\alpha_{SL}$), 
two-track invariant mass in the hypothesis of charged pions ($m_{\pi\pi}$), 
$\Delta p = |\vec{p}_{\rm sum}| - |\vec{p}_{K_S}|$.}  
 \label{fig:Variables}
\end{figure}

A multivariate analysis is performed with a boosted decision tree (BDT) classifier built with the following five variables with good discriminating power against background:
\begin{itemize}
\item[$p_1 , p_2$] : the tracks momenta;
\item[$\alpha_{1,2}$] : the angle at the vertex between the two momenta 
in the $K_S$ reference system;
\item[$\alpha_{LS}$] : the angle between the momentum sum, 
$\vec{p}_{\rm sum} = \vec{p}_1 + \vec{p}_2$, and the $K_L$--crash direction;
\item[$\Delta p$] : the difference between $|\vec{p}_{\rm sum}|$ and the absolute value
$|\vec{p}_{K_S}|$ of the $K_S$ 
momentum;
\item[$m_{\pi\pi}$] : the invariant mass reconstructed 
from $\vec{p}_1$ and $\vec{p}_2$, in the hypothesis of charged-pion mass. 
\end{itemize}
Figure~\ref{fig:Variables} shows the distributions of the variables
for data and simulated signal and background events. Two selection cuts are applied to avoid regions far away from the signal where MC does not reproduce well the data:
\begin{equation}
p < 320 {\rm \ MeV \ for \ both \ tracks}
\qquad {\rm and} \qquad
\Delta p < 190 {\rm \ MeV .}
\label{eq:ADDITIONALcut}
\end{equation}

\begin{figure}[htb!]
 \centering
 \includegraphics[width = 8.cm]{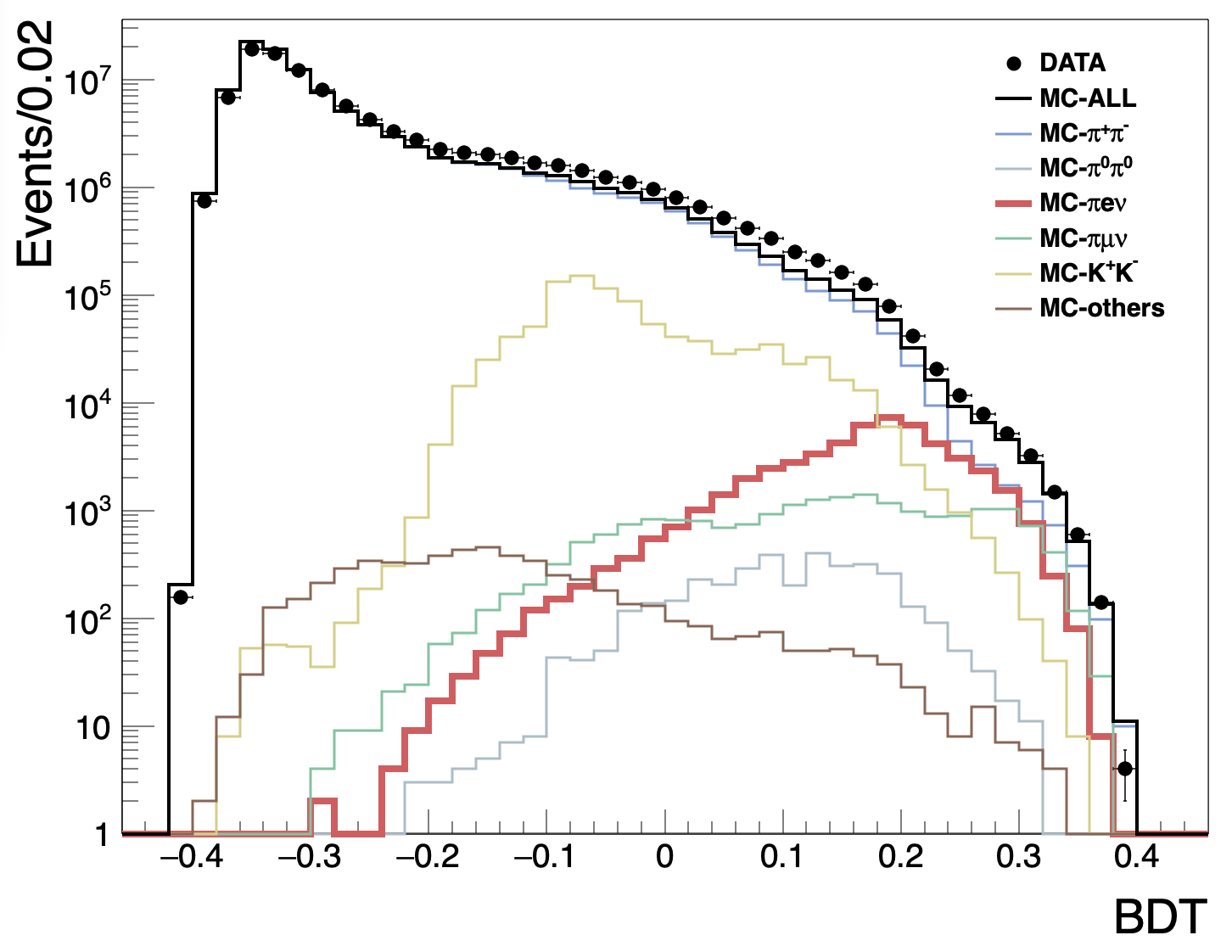}
 \caption{Distribution of the BDT classifier output for data and simulated signal and 
background events.}
 \label{fig:BDToutput}
\end{figure}

Training of BDT classifier is done with MC samples: 5,000 
$K_S \to \pi e \nu$ events and 50,000 background events. Samples of the same size are used for the test. 
After training and test the classification is run on both MC and data samples. Figure~\ref{fig:BDToutput} shows the BDT classifier output for data and simulated signal and background events. 
To suppress the large background contribution from $K_S \to \pi^+ \pi^-$ and $\phi \to K^+ K^-$ events, a cut is applied on the classifier output:
\begin{equation}
BDT > 0.15  .
\label{eq:BDTcut}
\end{equation}

\begin{figure}[htb!]
 \centering
 \includegraphics[width = 8.cm]{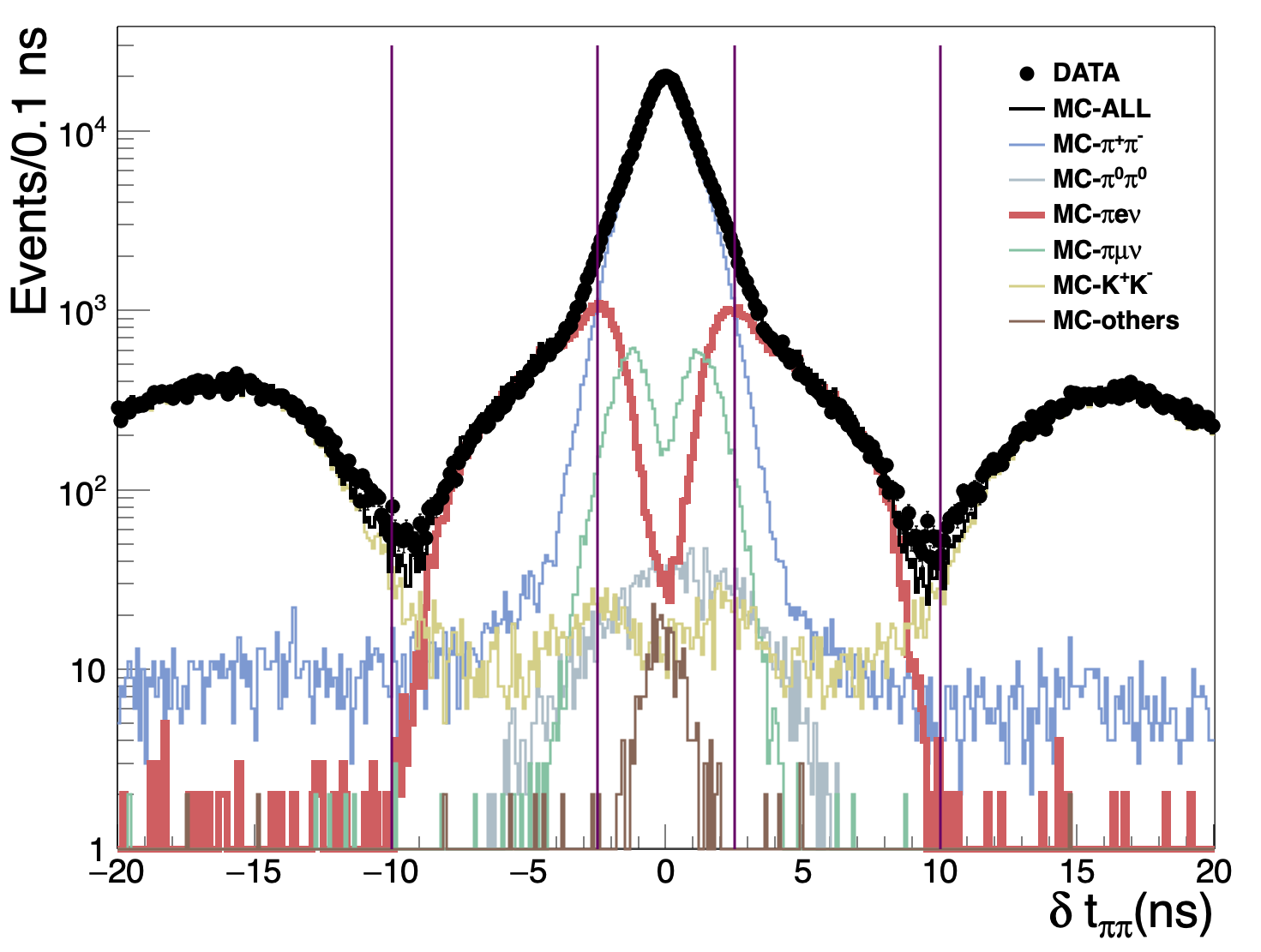}
 \caption{Distributions of $\delta t_{\pi\pi}$ for data and simulated signal and background events.
 The vertical lines indicate the selected range 2.5 ns $< |\delta t_{\pi \pi}| <$ 10 ns.}
 \label{fig:dTOFpipi}
\end{figure}

\begin{figure}[htb!]
 \centering
 \begin{tabular}{@{}cc@{}}
\includegraphics[width = 0.45\textwidth]{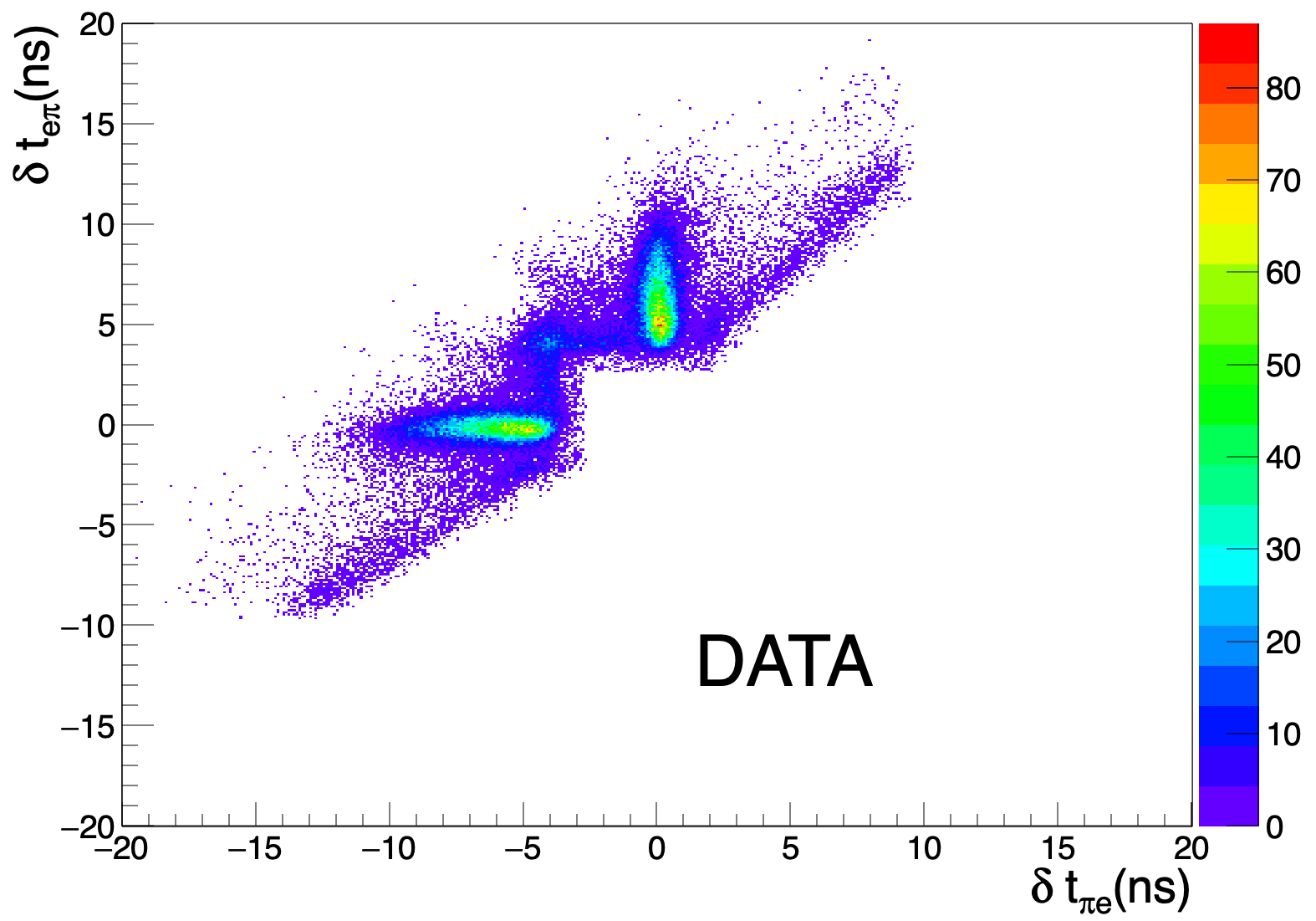}&
\includegraphics[width = 0.45\textwidth]{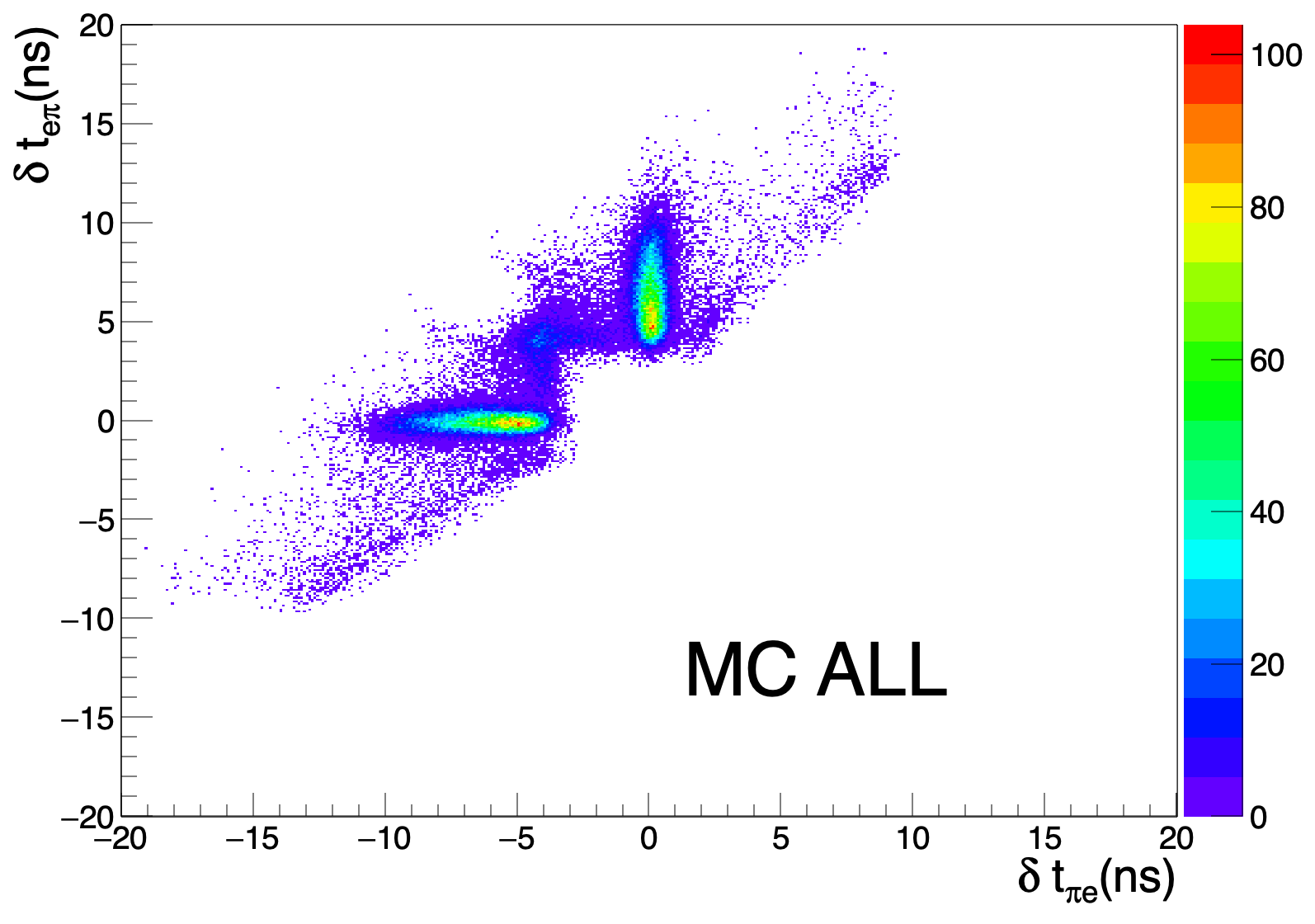}\\
\includegraphics[width = 0.45\textwidth]{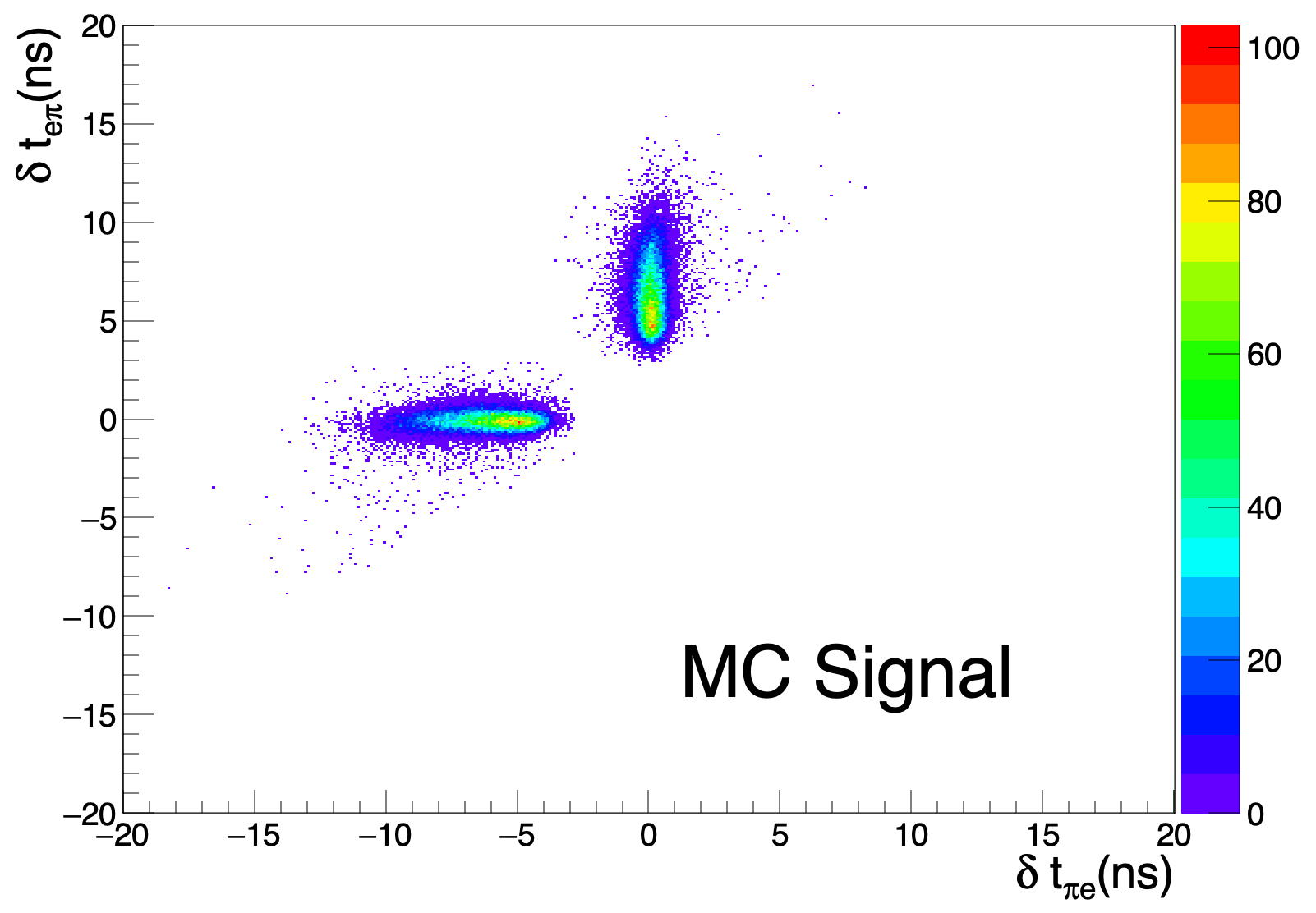}\\
  \end{tabular}
 \caption{Two dimensional distribution $(\delta t_{\pi e},\delta t_{e \pi})$ for top: data and MC all, bottom: MC signal.}
\label{fig:dTOFpie} 
\end{figure}

Track pairs in the selected events
are $e\pi$ for the signal and are $K\pi$, $\pi\pi$, $\mu\pi$ for the 
main backgrounds. A selection
based on time-of-flight measurements is performed to identify $e\pi$ pairs.
For each track associated to a cluster, the difference between the time-of-flight
measured by the calorimeter and the flight time measured 
along the particle trajectory
\begin{equation}
\delta t_i = t_{{\rm clu},i} - L_i / c \beta_i \qquad i = 1, 2
\label{eq:deltat_12}
\end{equation}
is computed, where $t_{{\rm clu},i}$ is the time associated to track $i$, $ L_i$ is the 
length of the track, and the velocity $\beta_i = p_i/\sqrt{p_i^2 + m_i^2}$ is function 
of the mass hypothesis for the particle with track $i$. 
The times $t_{{\rm clu},i}$ are referred to the trigger and the same T$_0$ value 
is assigned to both clusters. To reduce the uncertainty from the determination 
of T$_0$ the difference 
\[ \delta t_{1,2} = \delta t_1 - \delta t_2  \]
is used to determine the mass assignment.
The $\pi\pi$ hypothesis is tested first. Figure~\ref{fig:dTOFpipi} shows the
$\delta t_{\pi\pi} = \delta t_{1,\pi} - \delta t_{2, \pi}$ distribution. A fair agreement is observed 
between data and simulation,
with $K_S \to \pi e \nu$ and $K_S \to \pi \mu \nu$ distributions well separated
and large part of the $K^+ K^-$ background isolated in the tails of the distribution.
However the signal is hidden under a large $K_S \to \pi^+ \pi^-$ background, therefore a cut 
\begin{equation}
2.5\ {\rm ns} < |\delta t_{\pi\pi}| < 10\ {\rm ns}
\label{eq:deltaTOFpipicut}
\end{equation}
is applied.
Then, the $\pi e$ hypothesis 
is tested by assigning the pion and electron mass to either track defining
\[ \delta t_{\pi e} = \delta t_{1,\pi} - \delta t_{2,e}
\qquad {\rm and} \qquad
\delta t_{e \pi} = \delta t_{1,e} - \delta t_{2,\pi}, \]
where the label as track-1 and track-2 is chosen at random.  
Figure~\ref{fig:dTOFpie} shows the two-dimensional $(\delta t_{\pi e},\delta t_{e \pi})$ distribution for data and MC where signal events populate either band around $\delta t = 0$. 
The mass assignment is based on the comparison of 
two hypotheses: if 
$|\delta t_{1,\pi} - \delta t_{2,e}| < |\delta t_{1,e} - \delta t_{2,\pi}|$ 
track-1 is assigned to the pion and track-2 to the electron, 
otherwise the other solution is taken; the corresponding time difference, $\delta t_e$, is the 
value defined by $\texttt{min}[ | \delta t_{\pi e} |, | \delta t_{e \pi} | ]$.
A cut is applied on this variable 
\begin{equation}
| \delta t_e | < 1\ {\rm ns} .
\label{eq:deltaTOFpie}
\end{equation}

The number of events selected by the time-of-flight requirements is 57577 and the composition as predicted by simulation is listed in Table~\ref{tab:EventSel}. The background comprises $K_S \to \pi^+\pi^-$, $\phi \to K^+K^-$ and $K_S \to \pi \mu \nu$, the other contributions being small.
\begin{table}[htb]
\caption{Number of events after the BDT and TOF selections.}
\begin{center}
\begin{tabular}{lrr}
& Events & Fraction  [\%] \\ 
\hline
Data 	& 57\ 577            & \\
MC 		& 56\ 843            &  \\
\hline
$\phi \to K_L K_S, K_S \to \pi e \nu$     & 53\ 559 & 94.22 \\
$\phi \to K_L K_S, K_S \to \pi^+\pi^-$    & 2\ 175 & 3.83 \\
$\phi \to K^+K^-$        & 903     & 1.59 \\
$\phi \to K_L K_S, K_S \to \pi \mu \nu$  & 136    & 0.24  \\
others                          &   70 & 0.12 \\
\hline
\end{tabular}
\end{center}
\label{tab:EventSel}
\end{table}

The mass of the charged secondary identified as the electron is evaluated as
\[ m_e^2 = \left( E_{K_S} - E_{\pi} - p_{\rm miss} \right)^2 - p^2_e \]
with $p_{\rm miss}^2 = (\vec{p}_{K_S} - \vec{p}_{\pi} - \vec{p}_e)^2$,
$E_{K_S}$ and $\vec{p}_{K_S}$ being the energy  
and momentum reconstructed using the tagging $K_L$, and $\vec{p}_{\pi}$, 
$\vec{p}_e$, the momenta of the pion and electron tracks, respectively.

\begin{figure}[htb!]
\centering
\begin{tabular}{@{}cc@{}}
\includegraphics[width = 7.cm]{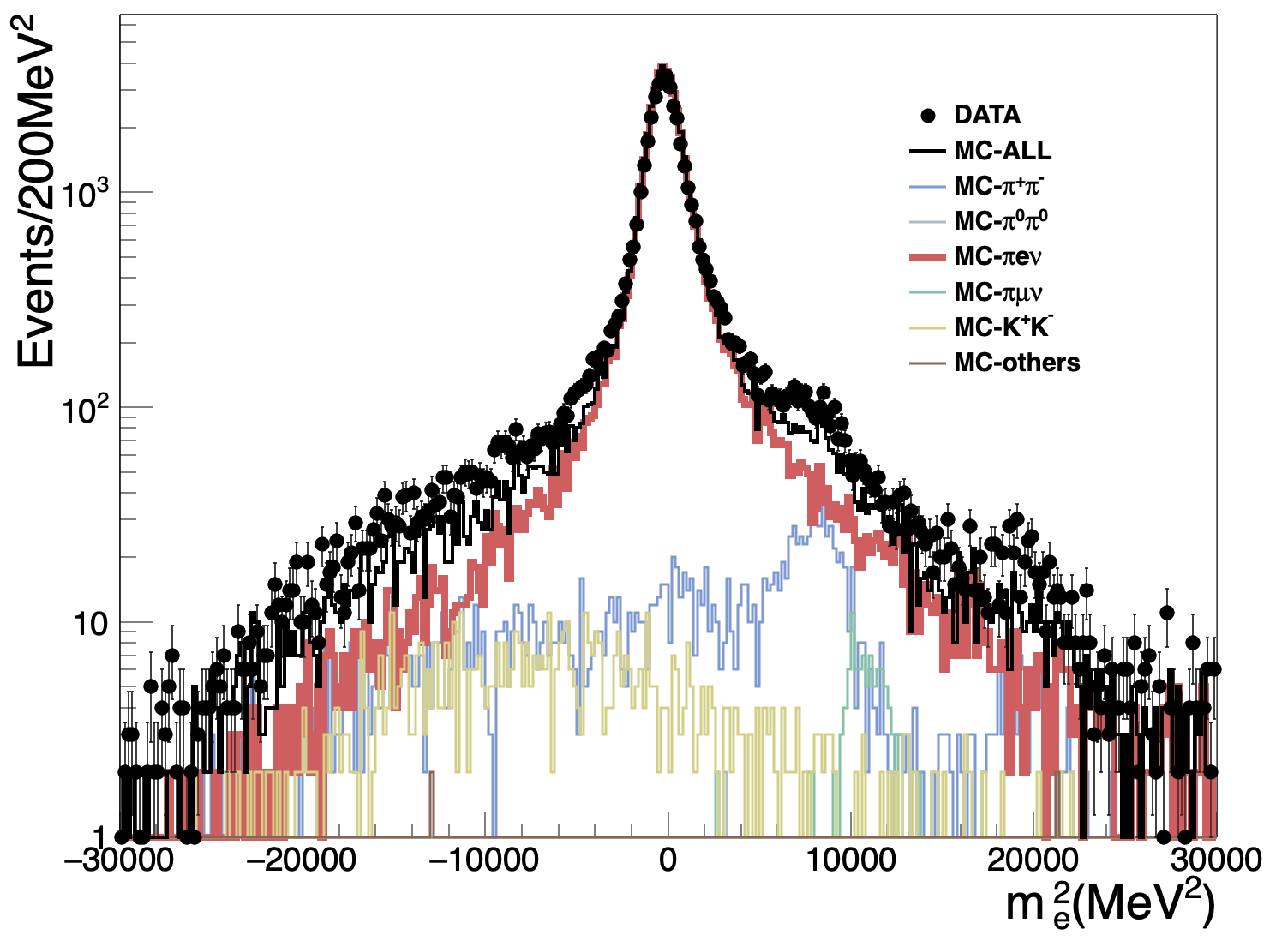}&
\includegraphics[width = 7.cm]{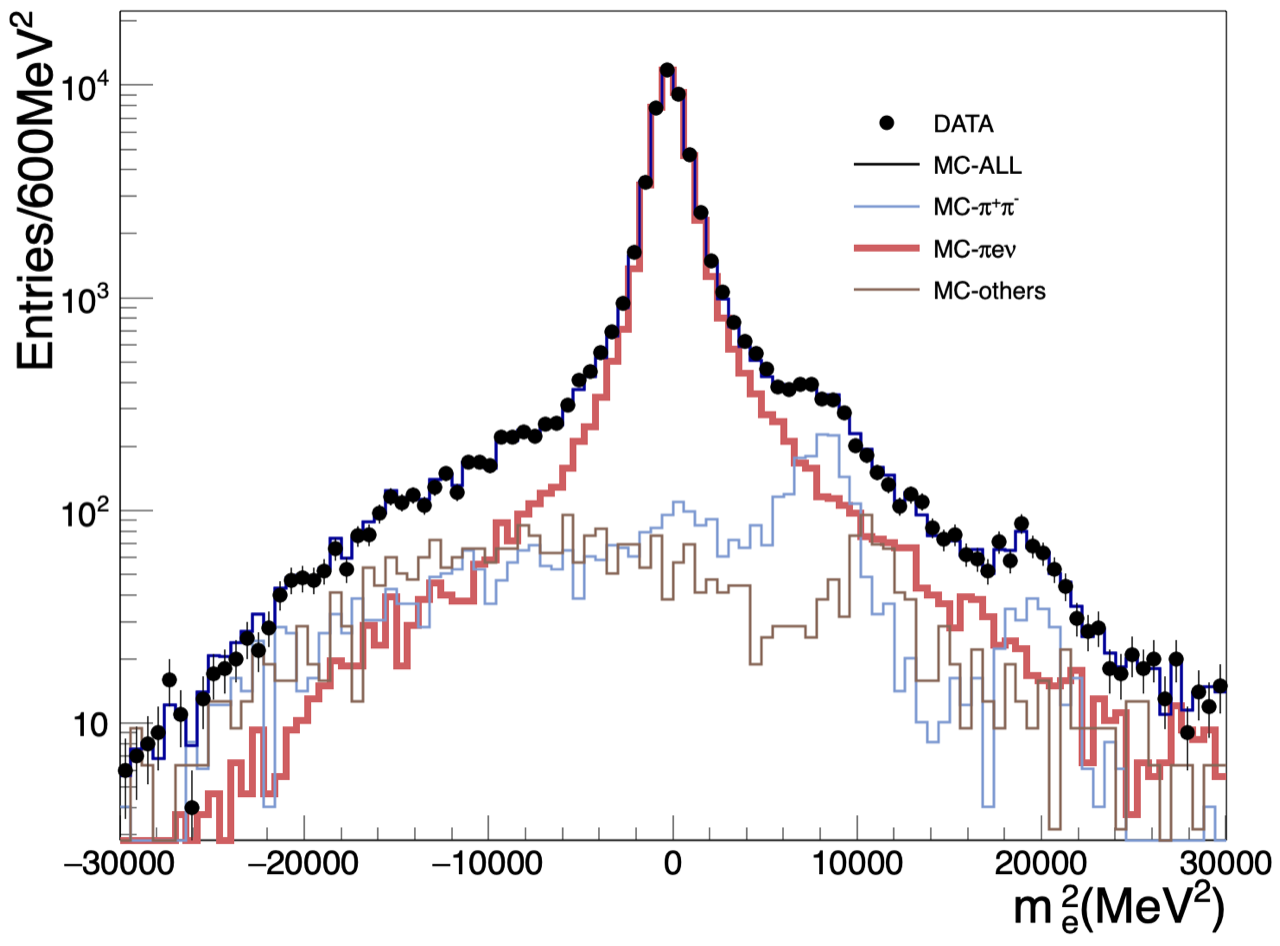} 
\end{tabular}
 \caption{The $m_e^2$ distribution for data, MC signal and background before the fit (left) and comparison of data with the result of the fit (right). }
 \label{fig:me2FIT}
\end{figure}

A fit to the $m_e^2$ distribution with the MC shapes of three components,
$K_S \to \pi e \nu$, $K_S \to \pi^+\pi^-$ and the sum of all other backgrounds, allows the number of signal events to be extracted. 
The fit is performed in 100 bins in the range [-30000,+30000] MeV$^2$. 
Figure~\ref{fig:me2FIT} shows the $m_e^2$ distribution
for data and simulated events before the fit, and the comparison of the fit output with the data. The fit result is reported in Table~\ref{tab:FitOutput}. 
The number of signal events is 
\[ N_{\pi e \nu} = 49647 \pm 316 \qquad {\rm with} \ \chi^2/{\rm ndf} = 76/96. \]

\begin{table}[htp]
\caption{Result of the fit to the $m_e^2$ distribution.}
\begin{center}
\begin{tabular}{lcc}
\hline
& Fraction & Events \\
\hline
$\pi e \nu $ & 0.8651 $\pm$ 0.0055 & 49\ 647 $\pm$ 316   \\
$\pi^+ \pi^- $ & 0.0763 $\pm$ 0.0068 & \ 4\ 379 $\pm$ 390    \\
all others & 0.0586 $\pm$ 0.0067 &    \ 3\ 363 $\pm$ 384 \\
\hline
Total &   & 57\ 389 \quad \qquad \ \\
\hline
$\chi^2$/{\rm ndf} & 76/96 \\
\hline
\end{tabular}
\end{center}
\label{tab:FitOutput}
\end{table}

The $K_S \to \pi^+ \pi^-$  normalisation sample is selected requiring 
$K_L$--crash, two opposite curvature tracks, the vertex as in Eq.~(\ref{eq:Vertex}) 
and $140 < p < 280$ MeV for both tracks (Figure~\ref{fig:Variables}(a)). 
A total of $N_{\pi \pi} = (282.314 \pm 0.017)\times10^6$ events are selected
with an efficiency of 97.4\% and a purity of 99.9\% as determined by simulation. 

\subsection{Determination of efficiencies} \label{EFFICIENCY}
The signal efficiency for a given selection is determined with a
$K_L \to \pi e \nu$ control sample (CS) and evaluated as
\begin{equation}
\epsilon_{\pi e \nu} = \epsilon_{\rm CS} \times 
\frac{\epsilon^{\rm MC}_{\pi e \nu}}{\epsilon^{\rm MC}_{\rm CS}} ,
\label{eq:EFFICIENCY}
\end{equation}
where $\epsilon_{\rm CS}$ is the efficiency of the control 
sample, and $\epsilon^{\rm MC}_{\pi e \nu}$,
$\epsilon^{\rm MC}_{\rm CS}$ are the efficiencies obtained from 
simulation for the signal and the control sample, respectively. 
Extensively studied with the KLOE detector~\cite{ref:KLtoany}, $K_L \to \pi e \nu$ decays are kinematically identical to the signal, the only difference being the much 
longer decay path.
Tagging is done with $K_S \to \pi^+ \pi^-$ decays
selected requiring two opposite curvature tracks and the vertex defined in Eq.~(\ref{eq:Vertex}) with the additional
requirement $|m_{\pi\pi} - m_{K^0}| < 15$ MeV to increase the purity, 
ensuring the angular and momentum resolutions are similar to the 
$K_L$--crash tagging for the signal.
The radial distance of the $K_L$ vertex is 
required to be smaller than 5 cm, to match the signal selection,
but greater than 1 cm to minimise the ambiguity in identifying
$K_L$ and $K_S$ vertices. 
Weighting the $K_L$ vertex position to emulate the $K_S$ vertex position has negligible effect on the result.

The control sample composition is $K_L \to \pi e \nu$ ($\mathcal{B} = 0.405$), $K_L \to \pi \mu \nu$ ($\mathcal{B} = 0.270$) and
$K_L \to \pi^+\pi^-\pi^0$ ($\mathcal{B} = 0.125$) decays, 
while most of $K_L \to \pi^0\pi^0 \pi^0$ 
decays are rejected requiring two tracks and the vertex.
The distribution of the $m^2_{\rm miss}$ missing mass,  with respect to the two tracks 
connected to the $K_L$ vertex and in the charged-pion mass hypothesis, shows a narrow isolated peak at the $\pi^0$ mass. $K_L \to \pi^+\pi^-\pi^0$ decays are efficiently rejected with the 
$m^2_{\rm miss} < 15000$ MeV$^2$ cut. 

Two control samples are selected, based on the two-step analysis strategy using largely uncorrelated variables and presented in Section~\ref{KtoPIENU}: the first CS$_{\rm kinBDT}$ applying a cut on the TOF variables to evaluate the efficiency of the selection based on the kinematic variables and the BDT classifier, the second CS$_{\rm TCATOF}$ applying a cut on kinematic variables to evaluate TCA and TOF selection efficiencies.

\begin{figure}[htb!]
 \centering
\includegraphics[width = 0.45\textwidth]{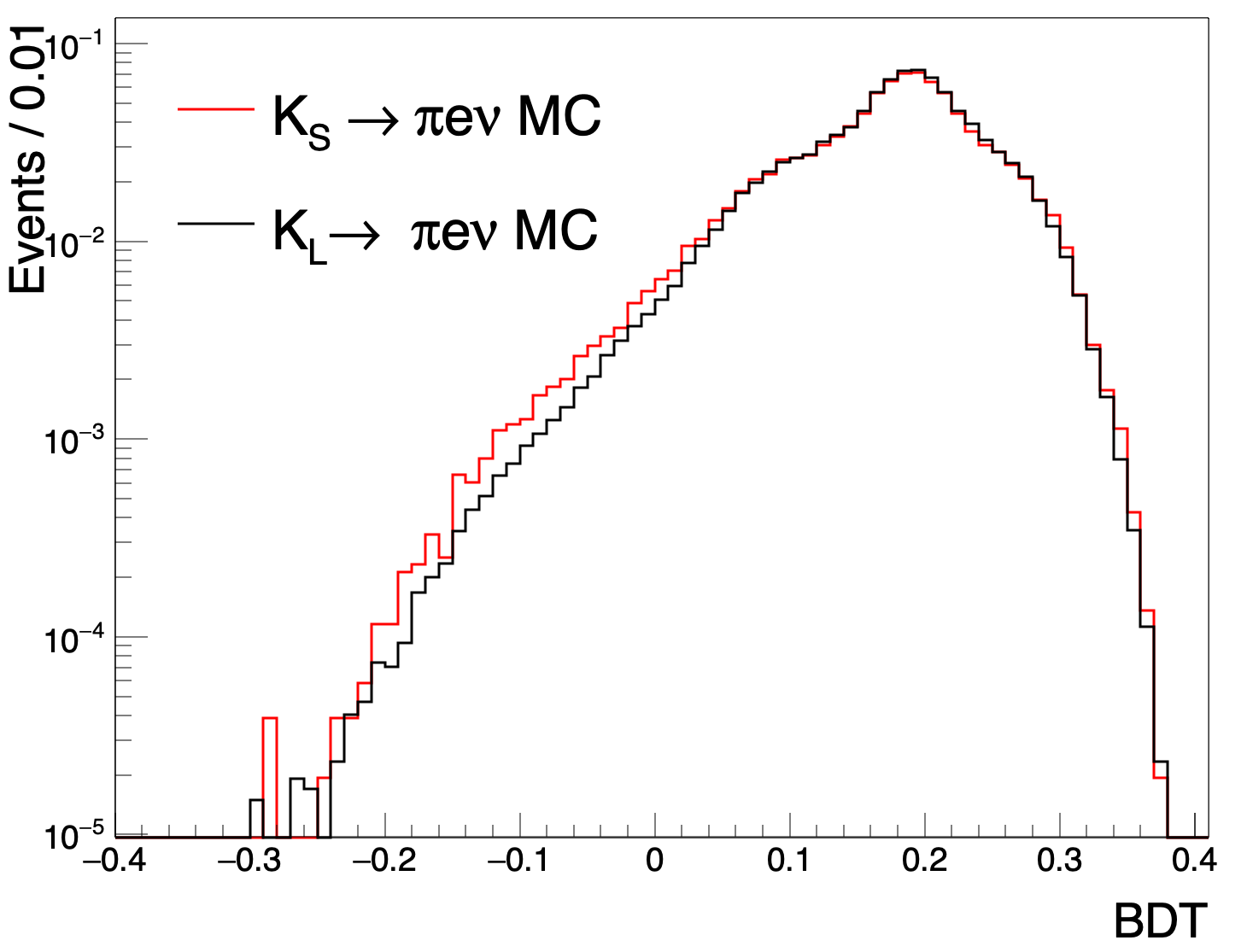}
\includegraphics[width = 0.45\textwidth]{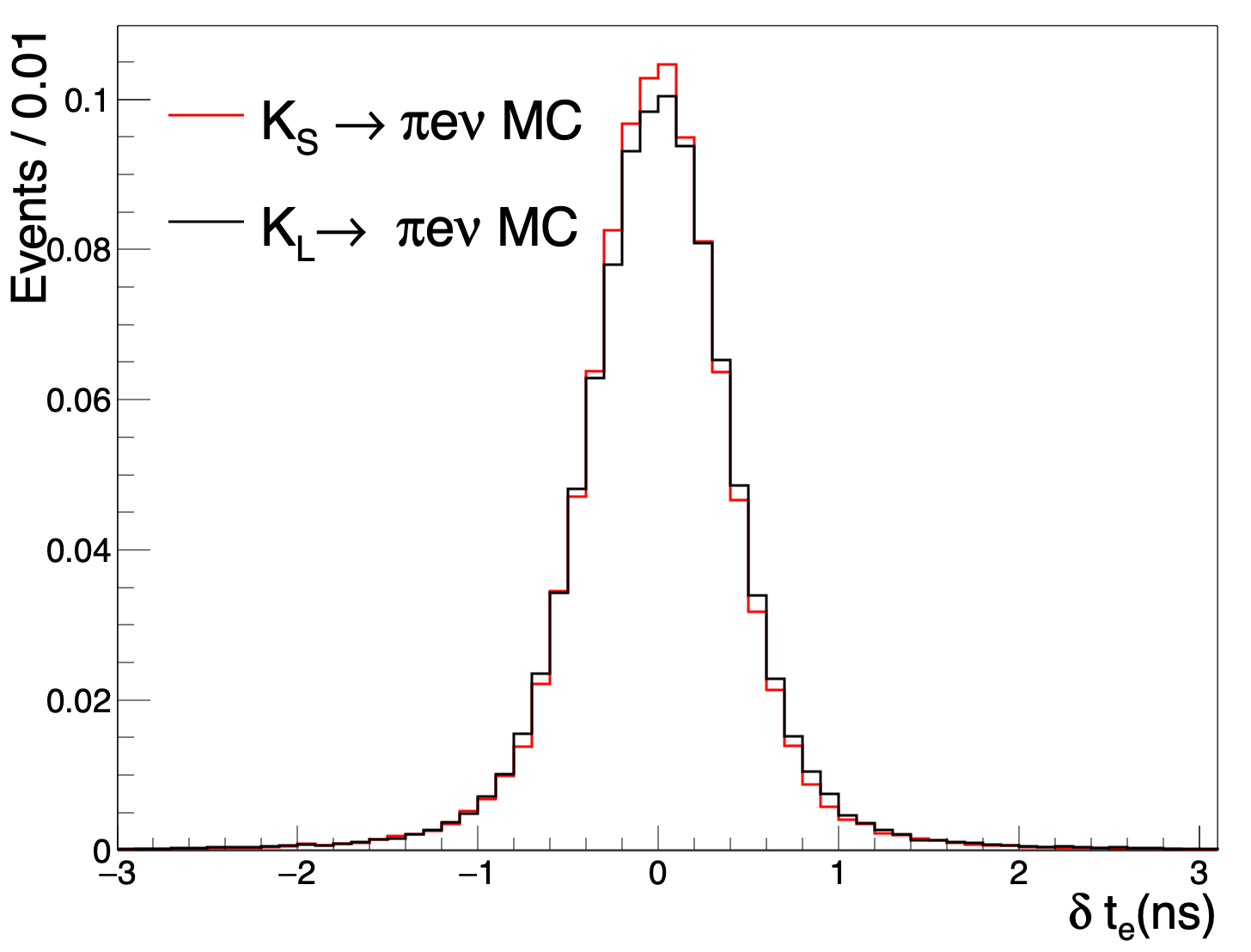}
 \caption{Monte Carlo distributions of the BDT classifier output (left) and
 $\delta t_e$ (right) for $K_L \to \pi e \nu$ (black) and $K_S \to \pi e \nu$ (red) events. }
 \label{fig:KLKSBdtTofMC}
\end{figure}

The CS$_{\rm kinBDT}$ control sample is selected applying a cut on the 
two-dimensional $(\delta t_{\pi e},\delta t_{e\pi})$
distribution, rejecting most of the $K_L \to \pi \mu \nu$ events.
The sample contains $0.44 \times 10^6$ events with a 97\% purity as determined from simulation.  
The Monte Carlo BDT distributions for the signal and control sample are compared 
in Figure~\ref{fig:KLKSBdtTofMC}(left). 
Applying to the control sample the same selections as for the signal,
Eqs.~(\ref{eq:ADDITIONALcut}) and~(\ref{eq:BDTcut}),
the efficiencies evaluated with Eq.~(\ref{eq:EFFICIENCY}) are
\[ \epsilon({\rm kin}) = 0.9720 \pm 0.0007_{\rm stat} \qquad {\rm and} 
\qquad \epsilon({\rm BDT}) = 0.6534 \pm 0.0013_{\rm stat} . \]

The CS$_{\rm TCATOF}$ control sample is selected applying a cut on the $(m_{\pi\pi},m^2_{\rm miss})$ 
distribution.
The sample contains $1.3 \times 10^6$ events with a 95\% purity as determined from simulation. 
In the $K_S \to \pi e \nu$ analysis, the T$_0$ is determined by the first 
cluster in time, associated with one of the tracks 
of the $K_S$ decay. Then, for the control sample the first cluster in time is required to be associated with the $K_L$ decay, in order not to
bias TOF variables. 
Figure~\ref{fig:KLKSBdtTofMC}(right) shows the comparison between the Monte Carlo distributions of $\delta t_e$ for signal and control sample.
Applying to the control sample the same selections as for the signal,
Eqs. (\ref{eq:deltaTOFpipicut}) anf (\ref{eq:deltaTOFpie}),
the efficiencies evaluated with Eq.~(\ref{eq:EFFICIENCY}) are
\[ \epsilon({\rm TCA}) = 0.4639 \pm 0.0009_{\rm stat} \qquad {\rm and} \qquad
\epsilon({\rm TOF}) = 0.6605 \pm 0.0012_{\rm stat} . \]
Table~\ref{tab:EffSum} summarises the signal selection efficiencies.
\begin{table}[htp!]
\caption{Signal selection efficiencies with statistical uncertainties. 
Correlations are accounted for in evaluating the total efficiency uncertainty.}
\begin{center}
\begin{tabular}{lc}
Selection & Efficiency \\ 
\hline
Preselection (from MC) & 0.9961 $\pm$ 0.0002 \\
Kin. variables selection & 0.9720 $\pm$ 0.0007 \\
BDT selection & 0.6534 $\pm$ 0.0013 \\
TCA selection & 0.4639 $\pm$ 0.0009  \\
TOF selection & 0.6605 $\pm$ 0.0012 \\
\hline
Total & 0.1938  $\pm$ 0.0006 \\
\end{tabular}
\end{center}
\label{tab:EffSum}
\end{table}

For the $K_S \to \pi^+\pi^-$ normalisation sample, the efficiency of 
the momentum selection $140 < p < 280$ MeV is determined using preselected data. The cut on the vertex transverse position in Eq.~(\ref{eq:Vertex}) is varied in 1 cm steps from $\rho^{\rm max}_{\rm vtx} = 1$ cm to 
$\rho^{\rm max}_{\rm vtx} = 4$ cm,
based on the observation that $\rho_{\rm vtx}$ and the tracks momenta are the least correlated variables, the correlation coefficient being 13\%. Using Eq.~(\ref{eq:EFFICIENCY}) and extrapolating to 
$\rho^{\rm max}_{\rm vtx} = 5$ cm, the efficiency is
$\epsilon^{\rm data}_{\pi\pi} = (96.569 \pm 0.004)\%$. 
Alternatively, the efficiency is evaluated using the $K_S \to \pi^+\pi^-$ data sample
(with $\rho^{\rm max}_{\rm vtx} = 5$ cm and 
$\epsilon^{\rm MC}_{\pi\pi} = \epsilon^{\rm MC}_{\rm pres}$), the efficiency is
$\epsilon^{\rm data}_{\pi\pi}  = (96.657 \pm 0.002)\%$. The second value, free from bias of variables correlation, is used for the efficiency and
the difference between the two values is taken as systematic uncertainty. 
The number of 
$K_S \to \pi^+ \pi^-$ events corrected for the efficiency is
$N_{\pi\pi}/\epsilon_{\pi\pi} = (292.08 \pm 0.27)\times10^6$.

The ratio $R_{\epsilon}$ in Eq.~(\ref{eq:RATIO}) includes
several effects depending on the event global properties: 
trigger, on-line filter, event classification, T$_0$ determination, 
$K_L$--crash and $K_S$ identification. In Table~\ref{tab:EffRatio}
the various contributions to $R_\epsilon$ evaluated with simulation 
are listed with statistical uncertainties only, the resulting value is 
$R_{\epsilon} = 1.1882 \pm 0.0017$. 
Systematic uncertainties are detailed in Section~\ref{SYSTEMATICS}.
\begin{table}[htb]
\caption{Ratios of MC efficiencies common to the $K_S \to \pi e \nu$ and $K_S \to \pi^+ \pi^-$ selections with statistical uncertainties. 
The error on $R_{\epsilon}$ is calculated as
the quadratic sum of the errors of the single ratios.}
\begin{center}
\begin{tabular}{lc}
Selection & $R_{\epsilon} = (\epsilon_{\pi\pi}/\epsilon_{\pi e \nu})_{\rm com}$ \\
\hline
Trigger & 1.0297 $\pm$ 0.0003 \\
On-line filter & 1.0054 $\pm$ 0.0001 \\
Event classification & 1.0635 $\pm$ 0.0004 \\
T0 time & 1.0063 $\pm$ 0.0001 \\
$K_L$--crash & 1.0295 $\pm$ 0.0010 \\
$K_S$ vertex reconstr. & 1.0418 $ \pm$ 0.0009 \\
\hline
$R_{\epsilon}$ & 1.1882 $\pm$ 0.0017 \\ 
\hline
\end{tabular}
\end{center}
\label{tab:EffRatio}
\end{table}

\section{Systematic uncertainties} \label{SYSTEMATICS}
The signal count is affected by three main systematic uncertainties: 
BDT selection, TOF selection, and the $m^2_e$ fit. 

The distributions of the BDT classifier
output for the data and simulated signal and control sample events are shown in Figures~\ref{fig:BDToutput} and~\ref{fig:KLKSBdtTofMC}.
The resolution of the BDT variable predicted by simulation comparing
the reconstructed events with those at generation level is
$\sigma_{\rm BDT} = 0.005$. The analysis is repeated varying the BDT cut 
in the range 0.135--0.17.
The ratio of the number of signal events determined with the $m_e^2$ fit
and the efficiency evaluated with Eq.~(\ref{eq:EFFICIENCY}) is found to be stable and the half-width of the band defined by the maximum and minimum values,
$\pm$0.27\%, is taken as relative systematic uncertainty.

The number of reconstructed clusters can be different for the signal ($K_L$--crash,$\pi e \nu$) and control sample ($\pi\pi$,$\pi e \nu$), thus the TCA efficiency calculation is repeated by weighting the events of the control sample by the number of track-associated clusters. The difference, less than 0.1\%, is taken as relative systematic uncertainty for the TCA efficiency.

The main source of uncertainty in the TOF selection is the lower cut on $|\delta t_{\pi\pi}|$ in Eq.~(\ref{eq:deltaTOFpipicut})
because the signal and background distributions in 
Figure~\ref{fig:dTOFpipi} are steep and with opposite slopes.
The resolution is the combination of the time resolution of the calorimeter, the tracking resolution of the drift chamber and the track-to-cluster association
and is determined by the width of the $\delta t_e$ 
distribution.

\begin{figure}[htb!]
 \centering
\includegraphics[width = 0.46\textwidth]{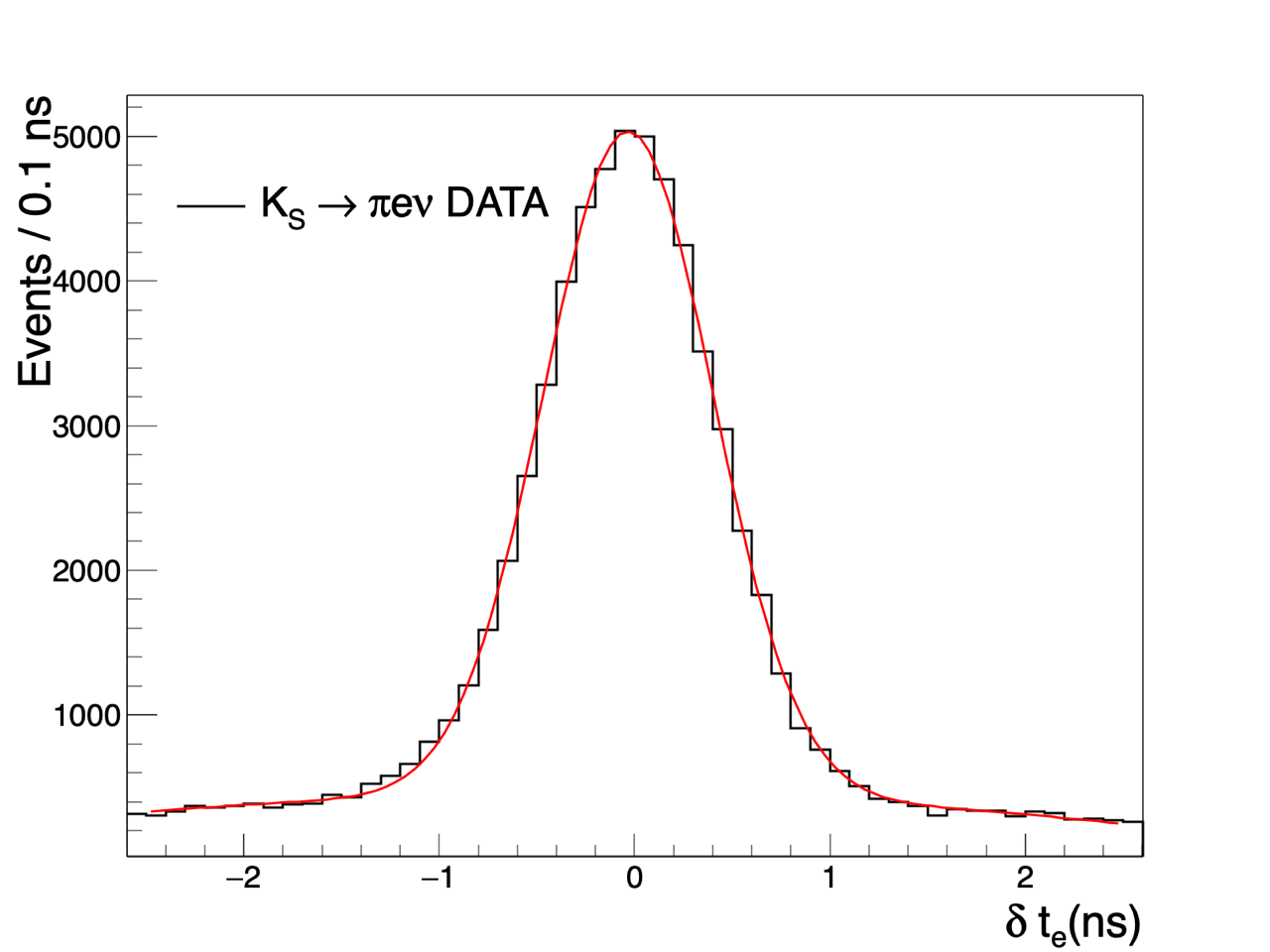}
 \includegraphics[width = 0.46\textwidth]{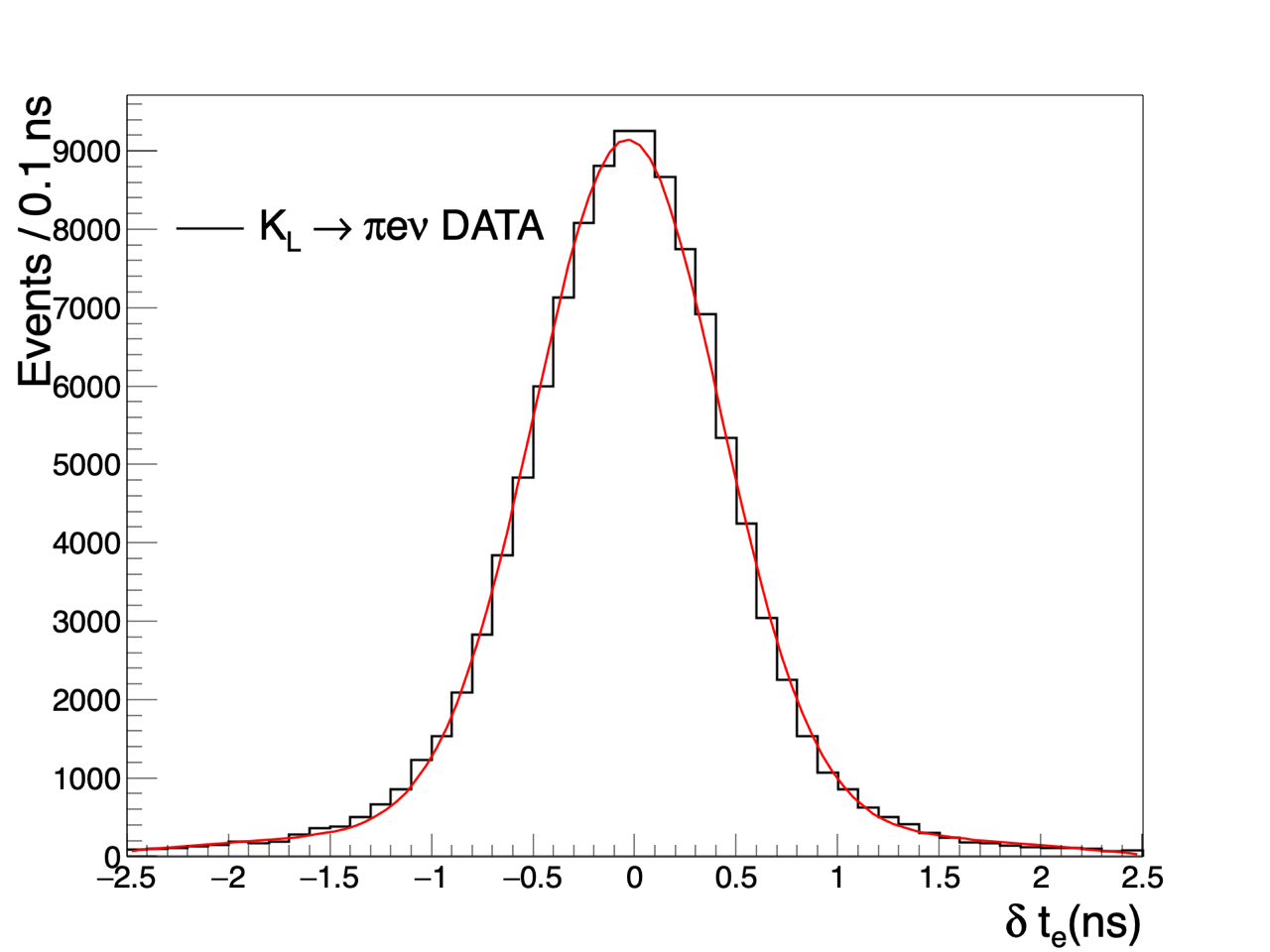}
 \caption{Comparison of the $\delta t_e$ distribution for the signal (left) and for the 
 $K_L \to \pi e \nu$ control sample (right).}
 \label{Ks-pienuDraft_deltat}
\end{figure}

The comparison of the $\delta t_e$ distributions for the signal and the
$K_L \to \pi e \nu$ control sample is shown in Figure~\ref{Ks-pienuDraft_deltat}, they are fitted with a Gaussian and a $2^{nd}$ degree polynomial, obtaining $\sigma = 0.44 \pm 0.02$ ns in both cases.
The analysis is repeated varying the $|\delta t_{\pi\pi}|$ lower cut in the range 2.0--3.0 ns, the half-width of the band gives a relative systematic uncertainty of 
$\pm$0.28\%.
With the same procedure the cut on $|\delta t_e|$ in  Eq.~(\ref{eq:deltaTOFpie}) is varied in the range 0.8--1.2 ns and the half-width of the band, $\pm$0.12\%, is taken as relative systematic uncertainty. 

Possible effects in the evaluation of the TCA and TOF efficiencies due to 
a
detector response different for the $\pi^+ e^- \bar{\nu}$ and $\pi^- e^+ \nu$ final states are negligible.

The fit to the $m_e^2$ distribution in Figure~\ref{fig:me2FIT}
is repeated varying the range and the bin size. The fit is also done using two separate components
for $K_S \to \pi \mu \nu$ and $\phi \to K^+K^-$, the $\chi^2$ is good but the statistical error is slightly increased. Half of the difference between maximum and minimum result of the different fits, $\pm 0.15\%$, is taken as relative systematic uncertainty.
The systematic uncertainties are listed in Table~\ref{tab:EffEvent}.
\begin{table}[htb]
\caption{Absolute systematic uncertainties.}
\begin{center}
\begin{tabular}{llcc}
& Selection & $\delta \epsilon_{\pi e \nu}^{\rm syst}$ [ $10^{-4}$ ] & $\delta \epsilon_{\pi^+\pi^-}^{\rm syst}$ [ $10^{-4}$ ] \\ 
\hline
                               & BDT selection & 5.3 & \\
$K_S \to \pi e \nu$ & TCA \& TOF selection & 6.0 & \\
                               & Fit parameters & 3.0 & \\
\hline
$K_S \to \pi^+ \pi^-$ & Event selection & & 8.8 \\
\hline
                               & Total & 8.5 & 8.8 \\
\hline
\end{tabular}
\end{center}
\label{tab:EffEvent}
\end{table}

The dependence of $R_{\epsilon}$ on systematic effects has been studied 
in previous analyses for different $K_S$ decays selected with the $K_L$--crash tagging method: 
$K_S \to \pi^+ \pi^-$ and $K_S \to \pi^0 \pi^0$~\cite{ref:KStopipi},
and $K_S \to \pi e \nu$~\cite{ref:KStopienuAsimmetry}.
The systematic uncertainties are evaluated by a comparison of data 
with simulation, the difference from one of the ratio $\frac{{\rm Data}}{{\rm MC}}$
is taken as systematic uncertainty. 

{\bf Trigger} -- Two triggers are used for recording the events, 
the calorimeter trigger and the drift chamber trigger. The validation of the 
MC relative efficiency is derived from the comparison of the single-trigger
and coincidence rates with the data. The data over MC ratio is 0.999 with 
negligible error.

{\bf On-line filter} -- The on-line filter rejects events triggered
by beam background, detector noise, and events surviving the
cosmic-ray veto. A fraction of non-filtered events prescaled 
by a factor of 20 allows to validate the MC efficiency of the filter. 
The data over MC ratio does not deviate from one by more than 0.1\%.

{\bf Event classification} -- The event classification produces
different streams for the analyses. The $K_L K_S$ stream used in this analysis selects events based on the properties of $K_S$ and $K_L$ decays.
In more than 99\% of the cases the events are selected based on the
$K_S$ decay topology and the $K_L$--crash signature and differences
between MC and data are accounted for in the systematic uncertainties 
described below for the $K_L$--crash and $K_S$ vertex reconstruction.

{\bf T$_0$} -- The trigger time is synchronised with the r.f. signal and the event T$_0$ is re-defined after event reconstruction.
The systematic uncertainty is evaluated analysing the 
data and MC distributions of T$_0$ for the decays with the
most different timing properties: 
$K_S \to \pi^+\pi^-$ and $K_S \to \pi^0\pi^0$~\cite{ref:KStopipi}.
The data over MC ratio does not deviate from one by more than 0.1\%.

{\bf $K_L$--crash} and {\bf $\beta^*$ selection.} -- The systematic uncertainty
is evaluated comparing data and simulated events tagged by 
$K_S \to \pi^+\pi^-$ and $K_S \to \pi^0\pi^0$ decays which have
different timing and topology characteristics. 
The data over MC ratio is 1.001 with negligible error.

{\bf $K_S$ vertex reconstruction} -- The systematic uncertainty of the requirement of 
two tracks forming a vertex in the cylinder defined by Eq.~(\ref{eq:Vertex}) is evaluated for signal and normalisation using a control sample of
$\phi \to \pi^+\pi^0\pi^-$ events selected requiring one track with minimum distance 
of approach to the beamline in the cylinder and a well-reconstructed $\pi^0$. 
Energy-momentum conservation determines the momentum of the second track. 
The momentum distribution of tracks in the control sample covers a range  wider than both signal and normalisation samples.
The efficiency for reconstructing the second track and the vertex is computed for data and simulation and the ratio $r(p_{\rm L},p_{\rm T}) = \frac{\epsilon^{\rm Data}}{\epsilon^{\rm MC}}$ is parameterised as function of the longitudinal and transverse momentum 
$p_{\rm L}$ and $p_{\rm T}$. 
The ratios relative to the signal and normalisation events, $r_{\pi e \nu}$ and 
$r_{\pi^+\pi^-}$, are obtained as convolution of $r(p_{\rm L},p_{\rm T})$ with the respective momentum distribution after preselection. The ratio  $\frac{r_{\pi^+\pi^-}}{r_{\pi e \nu}}$ deviates from one by $0.45\%$ with an uncertainty of 
0.2\% due to the knowledge of the parameters of the 
$r(p_{\rm L},p_{\rm T})$ function.

The $R_{\epsilon}$ total systematic uncertainty is estimated by combining the differences from one of the data over MC ratios and amounts to 0.48\%.
Including the systematic uncertainties the factors in Eq.~(\ref{eq:RATIO}) are:
\begin{equation}
\begin{array}{l}
\epsilon_{\pi^+\pi^-} = (96.657 \pm 0.088)\% , \\
\epsilon_{\pi e \nu} = (19.38 \pm 0.10)\% , \\
{\rm and} \quad R_{\epsilon} = 1.1882\pm 0.0059 .
\end{array}
\label{eq:systematics1}
\end{equation}

\section{The result} \label{RESULT}
Using Eq.~(\ref{eq:RATIO}) with $N_{\pi e \nu} = 49647 \pm 316$,
$N_{\pi\pi} = (282314 \pm 17)\times10^3$ events and the efficiencies
of Eq. (\ref{eq:systematics1}) we derive the ratio 
\[ \mathcal{R} =
(1.0421 \pm 0.0066_{\rm stat} \pm 0.0075_{\rm syst})\times10^{-3} . \]
The previous result from KLOE based on an independent data sample corresponding to an
integrated luminosity of 0.41 fb$^{-1}$ is 
$\mathcal{R} = (1.019 \pm 0.011_{\rm stat} \pm 0.007_{\rm syst})\times10^{-3}$~\cite{ref:KStopienu}.
Correlations exist between the two measurements in the determination of efficiencies 
for the event preselection and time-of-flight analysis, 
correlations in the determination of $R_{\epsilon}$ and the fit being negligible.
The correlation coefficient is 12\%. The combination of the two measurements
gives
\[ \mathcal{R} = \frac{\Gamma(K_S \to \pi e \nu)}{\Gamma(K_S \to \pi^+ \pi^-)} = 
(1.0338 \pm 0.0054_{\rm stat} \pm 0.0064_{\rm syst})\times10^{-3} . \]
Using the value 
$\mathcal{B}(K_S \to \pi^+ \pi^-) = 0.69196 \pm 0.00051$ 
measured by KLOE~\cite{ref:KStopipi}, we derive the branching 
fraction 
\[ \mathcal{B}(K_S \to \pi e \nu)
= (7.153 \pm 0.037_{\rm stat} \pm 0.044_{\rm syst} )\times10^{-4}
= (7.153 \pm 0.058)\times10^{-4} . \]
The value of $|V_{us}|$ is related to the $K_S$ semileptonic branching fraction by the equation
\[ \mathcal{B}(K_S \to \pi \ell \nu) = 
\frac{G^2 (f_+(0) |V_{us}|)^2}{192 \pi^3} \tau_S m_K^5 
I^{\ell}_K S_{\rm EW} (1 + \delta^{K\ell}_{\rm EM}) , \] 
where $I^{\ell}_K$ is the phase-space integral, which depends on measured semileptonic
form factors, $S_{\rm EW}$ is the short-distance electro-weak correction, 
$\delta^{K\ell}_{\rm EM}$ is the mode-dependent long-distance radiative correction, and $f_+(0)$ is the form factor at zero momentum transfer for the $\ell \nu$ system. 
Using the values $S_{\rm EW} = 1.0232 \pm 0.0003$~\cite{ref:Sirlin1993}, 
$I^e_K = 0.15470 \pm 0.00015$ and 
$\delta^{Ke}_{\rm EM} = (1.16 \pm 0.03)\ 10^{-2}$ from 
Ref.~\cite{ref:VusUpdate}, and the world average values for the $K_S$ mass and lifetime~\cite{ref:PDG}  we derive
\[ f_+(0) |V_{us}| = 0.2170 \pm 0.0009 .  \]

\section{Conclusion} 
A measurement of the ratio $\mathcal{R} = \Gamma(K_S \to \pi e \nu) / \Gamma(K_S \to \pi^+ \pi^-)$
is presented based on data collected with the KLOE experiment at the DA$\Phi$NE $\phi$-factory 
corresponding to an integrated luminosity of 1.63 fb$^{-1}$.
The $\phi \to K_LK_S$ decays are exploited to select samples of pure and quasi-monochromatic
$K_S$ mesons and data control samples of $K_L \to \pi e \nu$ decays. 
The $K_S$ decays are tagged by the detection of a $K_L$ interaction in the 
detector. The $K_S \to \pi e \nu$ events are selected by
a boosted decision tree built with kinematic variables and by measurements of time-of-flight.
The efficiencies for detecting the $K_S \to \pi e \nu$ decays are derived from 
$K_L\to \pi e \nu$ data control samples. 
A fit to the $m^2$ distribution of the identified electron track
finds $49647 \pm 316$ signal events. Normalising to $K_S \to \pi^+ \pi^-$ decay events recorded in the same dataset, the result 
is $\mathcal{R} =
(1.0421 \pm 0.0066_{\rm stat} \pm 0.0075_{\rm syst})\times10^{-3}$.
The combination with our previous measurement 
gives $\mathcal{R} =
(1.0338 \pm 0.0054_{\rm stat} \pm 0.0064_{\rm syst})\times10^{-3}$.
From this value we derive the branching fraction
 $\mathcal{B}(K_S \to \pi e \nu) = (7.153 \pm 0.037_{\rm stat} \pm 0.044_{\rm syst} )\times10^{-4}$ and the value of $|V_{us}|$ times the form factor at zero momentum transfer, $f_+(0) |V_{us}| = 0.2170 \pm 0.009$.

\acknowledgments
We warmly thank our former KLOE colleagues for the access to the data collected during the KLOE data taking campaign.
We thank the DA$\Phi$NE team for their efforts in maintaining good running conditions and their collaboration during both the KLOE run 
and the KLOE-2 data taking with an upgraded collision scheme \cite{DAPHNE2, DAPHNE3}.
We are very grateful to our colleague G. Capon for his enlightening comments and suggestions about the manuscript.
We want to thank our technical staff: 
G.F. Fortugno and F. Sborzacchi for their dedication in ensuring efficient operation of the KLOE computing facilities; 
M. Anelli for his continuous attention to the gas system and detector safety; 
A. Balla, M. Gatta, G. Corradi and G. Papalino for electronics maintenance; 
C. Piscitelli for his help during major maintenance periods. 
This work was supported in part 
by the Polish National Science Centre through the Grants No.
2014/14/E/ST2/00262,
2016/21/N/ST2/01727,
2017/26/M/ST2/00697.

\end{document}